\documentstyle[epsfig]{article}

\begin{document}


\begin{center}
{\bf{{\bf apeNEXT}: A MULTI-TFLOPS LQCD COMPUTING PROJECT}}
\vspace*{0.5cm}

{\bf{R.~Alfieri, R.~Di~Renzo, E. Onofri}}\\
Dipartimento di Fisica, Universit\`a di Parma, and\\
INFN, Gruppo collegato di Parma, Parco Area delle Scienze, I-43100 Parma,
Italy.
\vspace{0.3cm}

{\bf{A.~Bartoloni, C.~Battista, N.~Cabibbo, M.~Cosimi,}\\
\bf{ A.~Lonardo, A.~Michelotti, F.~Rapuano, B.~Proietti,}\\ 
\bf{D.~Rossetti, G.~Sacco, S. Tassa,}\\
\bf{M.~Torelli, P.~Vicini}}\\
Dipartimento di Fisica, Universit\`a di Roma \lq La Sapienza\rq~and\\ 
INFN, Sezione di Roma, P.le A. Moro 2, I-00185 Roma, Italy.
\vspace{0.3cm}

{\bf{Ph.~Boucaud, O.~P\`ene}}\\
Laboratoire de Physique Th\'eorique \\
Universit\'e de Paris-sud (Orsay), France.
\vspace{0.3cm}

{\bf{W.~Errico, G.~Magazz\`u, L.~Sartori, F.~Schifano, R.~Tripiccione}}\\
INFN, Sezione di Pisa, Via Livornese 1291, I-56010 San Piero a Grado, Italy.
\vspace{0.3cm}

{\bf{P.~De Riso, R.~Petronzio}}\\
Dipartimento di Fisica, Universit\`a di Roma II \lq Tor Vergata\rq~and\\
INFN, Sezione di Roma II, Via della Ricerca Scientifica, 1 - 00133 Roma,Italy.
\vspace{0.3cm}

{\bf{C.~Destri, R.~Frezzotti, G.~Marchesini}}\\
Dipartimento di Fisica, Universit\`a di Milano-Bicocca and\\
INFN, Sezione di Milano, Via Celoria 16, I-20100 Milano, Italy.
\vspace{0.3cm}

{\bf{U.~Gensch, A.~Kretzschmann, H.~Leich,}\\
\bf{N.~Paschedag, U.~Schwendicke, H.~Simma,}\\
\bf{R.~Sommer, K.~Sulanke, P.~Wegner}}\\
DESY, Platanenallee 6, D-15738 Zeuthen, Germany.
\vspace{0.3cm}

{\bf{D. Pleiter, K. Jansen}}\\
NIC@DESY, Platanenallee 6, D-15738 Zeuthen, Germany.
\vspace{0.3cm}

{\bf{A.~Fucci, B.~Martin, J.~Pech}}\\
CERN, CH-1211 Geneva 23, Switzerland.
\vspace{0.2cm}
\eject
              
{\bf{E.~Panizzi}}\\
Dipartimento di Ingegneria Elettrica, Universit\`a de l'Aquila and \\
INFN, Sezione di Roma, P.le A. Moro 2, I-00185 Roma, Italy.
\vspace{0.3cm}

{\bf{A.~Petricola}}\\
Dipartimento di Ingegneria Elettrica, Universit\`a de l'Aquila and \\
INFN, Laboratori Nazionali del Gran Sasso, Assergi, Italy.

\eject
\vspace*{2cm}
\noindent
{\bf ABSTRACT} \\ \end{center}
\vspace*{5mm}

This paper is a slightly modified and reduced version of the proposal of the
{\bf apeNEXT} project, which was submitted to DESY and INFN in spring 2000.
It presents the basic motivations and ideas of a next generation lattice
QCD (LQCD) computing project, whose goal is the construction and operation of
several large scale Multi-TFlops LQCD engines, providing an integrated peak
performance of tens of TFlops, and a sustained (double precision)
performance on key LQCD kernels of about 50\% of peak speed. The software
environment supporting these machine is organized in such a way that it allows
relatively easy migration between {\bf apeNEXT} and more traditional computer
systems. We describe the physics motivations behind the project and the
hardware and software architecture of the new LQCD engine.
\eject

\section{Introduction}
Several research groups in the Lattice QCD (LQCD) community  have
developed LQCD optimized massively parallel processors \cite{review}.
These systems have provided in the last decade a significant fraction of all
compute cycles available all over the world for lattice simulations. In this
framework, INFN and DESY have developed the APEmille parallel processor.
APEmille is a LQCD oriented massively parallel number-cruncher
\cite{a1000}, providing peak performance of several hundred Gflops.  The
first APEmille systems have been commissioned in late 1999. More machines
have become available since then and yet a few more will be built
in the near future (see later for details).

We expect APEmille machines to become the work-horse for LQCD computing in
several laboratories in Europe in the next few years. It is however
clear (and explained in detail in a following section) that APEmille is unable
to support serious LQCD simulations at the level expected after the year 2003. 

The continuing physics motivation to pursue numerical studies of lattice QCD
and the level of needed computing resources have been analyzed in detail by a
review panel appointed by the European Committee for Future Accelerator (ECFA)
\cite{ecfa}. We fully endorse the conclusions of the ECFA report (which can be
regarded as an ideal introduction to the present document). 
In this paper we present a proposal for a new lattice QCD project that
builds on the experience of the previous generation APE machines and tries to
implement several of the recommendations of the ECFA panel. This paper is an
enlarged and improved version of a preliminary proposal
\cite{oldprop}, submitted to the INFN Board of Directors in summer 1999.

The new project (that we refer to as {\bf apeNEXT}) is characterized by the 
following architectural goals:
\begin{itemize}
\item an expected peak performance for large machines in excess of 5 TFlops,
using double precision floating point arithmetics.
\item a sustained (double precision)
efficiency of about 50\% on key LQCD kernels
(such as the inversion of the Dirac operator). 
\item a large on-line data storage (512 GByte to 1 TByte for large machines).
\item input/output channels able to sustain a data-rate of
$0.5$ MByte/sec/GFlops.
\item a programming environment that allows relatively straightforward and
easy migration of physics codes between {\bf apeNEXT} and more traditional
computer systems.
\end{itemize}

From the point of view of the organization of the project, the following
points are in order:
\begin{itemize}
\item the {\bf apeNEXT} architecture will be very closely optimized to LQCD
simulations. In other words, {\bf apeNEXT} will be more tuned towards LQCD than 
APEmille.
\item The general know-how of APEmille, as well as several important building
blocks, will be heavily reused in the new project (properly rescaled to keep
technology advances into account). This is a key point that we plan to leverage
on, in order to shorten development time.
\item We plan from the beginning the installation of several large machines at
approximately the same time at several collaboration sites.
(Collaboration membership is also somewhat enlarged in comparison with
APEmille). Stated otherwise, we plan to build up very high processing
performance for LQCD (of the order of several tens of TFlops) by operating in a
loosely coordinated way several machines.
\item Provisions to facilitate an industrial exploitation of the project are
not one of the stated goals of the project. We do see however that
several building blocks of the project (most notably in the area of inter-node
communications) may have an important impact on other areas of computing for
physics (and, more generally, for cluster computing or farming).
We will do our best to make our results reusable.
\end{itemize}

This paper describes the hardware and software architecture that we plan to
develop. It does not cover the organization of the project, the proposed
schedule of our activities and any financial issues. These points are considered
elsewhere. The paper is organized as follows:
\begin{itemize}

\item{{\bf Section 2}} discusses the physics goals of the project and their
corresponding computing requirements (in terms of processing performance, data
storage, bandwidth).  
\item{{\bf Section 3}} briefly summarizes the APEmille architecture and
substantiates the need for a new project.
\item{{\bf Section 4}} reviews similar planned or started projects.
\item{{\bf Section 5}} surveys the status and prospects of some enabling
technologies for our project.
\item{{\bf Section 6}} discusses advantages and disadvantages of custom versus
off-the-shelf technologies for the processing element of the new computer. 
\item{{\bf Section 7}} presents the global architecture of our new
massively parallel LQCD machine.
\item{{\bf Section 8}} describes the details of the processing node.
\item{{\bf Section 9}} covers the architecture of the interconnection network.
\item{{\bf Section 10}} discusses several possible options for
the topology and the mechanical set-up of the system.
\item{{\bf Section 11}} is the first section on software. Here
we describe the programming environment that we plan to develop for
{\bf apeNEXT}.
\item{{\bf Section 12}} is a matching section discussing the operating system
and other system-software issues.
\item{{\bf Section 13}} reviews the design method that we plan to
follow in the development of the system.
\item{{\bf Section 14}} contains our conclusions.
\end{itemize}

\section{Physics Requirements}
In the definition of the new project we keep a clear focus on a 
very limited number of important physics simulation areas, that set
the requirements for the new project.

The translation of
physics requirements into machine parameters requires certain assumptions about 
the algorithms to be used. We base our considerations on tested algorithms
such as SSOR-preconditioned BiCGstab and Hybrid Monte Carlo, for Wilson
fermions with improved action \cite{algoreview}. New theoretical developments
(domain wall fermions, Wilson-Dirac operators satisfying the Ginsparg-Wilson 
relation, etc.) are likely to be implemented in a way which has  
computational characteristics very similar to the standard Dirac operator. 

We expect that in the years 2003-2006, large {\em production} 
LQCD simulations will be mainly focused on the following lines:

\begin{itemize}
\item full QCD simulations (including dynamical fermions) on
lattices with sizes of the order of $48^3 \times 96$ (a physical system
of $L = 2 \div 4 $ fm and $a = 0.1 \div 0.05 $ fm). Dynamical quark
masses should also decrease, with a reasonable target
corresponding to $m_\pi/m_\rho \simeq 0.35$ (although it is not
realistic to expect that both goals are obtained in the same simulation).
\item simulations in the quenched approximation on very large lattices
($100^3 \times 100 \div 200$) and large $\beta$
($L = 1.5 \div 2.0 $ fm and $a = 0.1 \div 0.02 $ fm) for the study of
$b$ physics with as little extrapolation as possible
in the mass of the heavy quark.
\end{itemize}

The first item is heavily CPU limited, since one has to solve 
the Dirac equation repeatedly during the updating process.
The second item is basically memory
limited, due to very large lattice size. In both cases, our
target is a resolution about two-times better than currently
possible (implying, as discussed later on, an increase in computing power
of two orders of magnitude).

As a guideline to define a new LQCD engine for these classes of problems, we
require that:
\begin{enumerate}
\item
The node-topology and communication network is
optimized for the lattice sizes required in full QCD simulations.
Since for many problems of LQCD it is important to perform
a finite-size scaling analysis, it is desirable that the machine
performs efficiently not
only on {\em large} but also on comparatively {\em small}
lattices, e.g., in full QCD  one may think of $N_{L}^3 \times N_{T}$
lattices with $N_{L} = 16, 20,...,32$ and 48, and $N_L \leq N_T \leq 2 N_L$.
For smaller lattices, as the required computing performance decreases,
more traditional machines (such as PC clusters) or previous
generation dedicated systems can be used.

\item
The communication network has enough bandwidth to handle the  large
degree of data exchange between neighbouring sites (and hence  
compute nodes) needed in
LQCD computations. The interconnect architecture should support the  natural
(APE-like) programming model with direct remote data access
\cite{taomanual}. This
approach minimizes software and memory overhead (and coding effort) for  
pre-loading of remote data.

\item
The processing nodes sustain high performance on the execution of the
arithmetic and control operations which are relevant for the codes 
(or at least their basic kernels) of full-QCD algorithms, in particular 
double precision floating point arithmetics, memory access to field 
variables of composed data structures, local and global program-flow 
control, etc.

To obtain a good floating-point efficiency for the execution
of a given computation, the compute power and memory bandwidth
should be balanced accordingly.
This balance is usually measured in term of the parameter $R$, defined
as the ratio between the number of floating-point operations and
the corresponding memory accesses (in the corresponding
data format). A processor is balanced for a given algorithm
if the $R$ value required by the algorithm is roughly equal to the 
$R$ value allowed by the processor itself. 
 In the case of the Dirac operator, 
which usually dominates the cost in LQCD computations, a typical 
value is $R \simeq4$.

\item
Memory size, disk space and disk-bandwidth match each other
and are well suited to the problems we
want to study. This means that all compute intensive kernels
must not be slowed significantly because required data is not available in main
memory. We must keep all data in physical memory as long
as possible. In all cases in which this is not possible (e.g., for
light-fermion propagators on very large lattices)
we must be able to temporarily store on (and retrieve from) disk
with large enough bandwidth.
\end{enumerate}

These requirements shape the global architecture of the machine:

\begin{enumerate}
\item
We consider architectures based on three dimensional grids of
processors, with nearest neighbour 
data-links. Reasonable sizes of the mesh of {\em processors} that
will be used for the simulation of large lattices are
somewhere
in the range $8^3 \cdots 12^3 \cdots 16^3$ nodes, where a physical lattice of
$48^3 \times 96$ points can be readily mapped. For finite size analysis
on small lattices, a mesh of $4^3 \cdots 6^3$ processors may be considered.

The size of the processor mesh dictates a lower bound on the communication
bandwidth between neighbouring processors. We define by $\rho$ the ratio of
local memory accesses (transfers between processor and its memory) over
remote memory accesses (transfers between neighbour processors),
which depends on the lattice size and the algorithm. Under the
assumption of balanced local bandwidth (i.e., processors are able to access
enough data in local memory to sustain their potential performance,
see later for details),
effective bandwidth\footnote{
 including the effect of the start-up latency for typical packet lengths.
}
for remote communications must not be lower than $1/\rho$ times the local
bandwidth.
Estimates of the required ratio
for a naive implementation of the Dirac operator using Wilson fermions
are given in table \ref{tab:table_remote} for a sub-lattice of 
$n_L^3\times N_T$ physical points and local time direction per processor
(note that, to first approximation, $\rho \simeq 2n_L$).

\begin{table}[hbt]\centering
\begin{tabular}{|c|c|}
\hline\hline
Linear lattice size & $\rho$ \\
\hline
$3^3$ & $5.8$\\
$4^3$ & $7.8$\\
$6^3$ & $11.6$\\
$8^3$ & $15.5$\\
\hline\hline
\end{tabular}
\vspace*{5mm}
\caption{Local vs. remote memory access patterns: $\rho$ is the
ratio of memory accesses to local memory over memory accesses to
neighbour nodes in a simple implementation of the solver for the
Dirac operator. $\rho$ is estimated as a function of the linear
size of the sub-lattice mapped onto {\em each} processor.}
\label{tab:table_remote}
\end{table}

A nice and simple trick can be used in the computation of the Dirac operator
to reduce the number of remote accesses. For the negative directions the 
Dirac operator involves terms of the type $U^\dag_\mu(x-\mu) \psi(x-\mu)$ where
the fermion term $\psi$ and the corresponding gauge matrix ($U$) must 
be fetched from the same place. We can therefore evaluate the product 
$U_\mu(x-\mu) \psi(x-\mu)$ on the remote node and transfer the
result only. In brief, all remote accesses involving gauge fields disappear.
Table \ref{tab:table_remote2} contains the $\rho$ values corresponding to the
evaluation of the Dirac operator using the above mentioned technique.
We consider the comfortably increased $\rho$ values as an useful safety
margin, that could be exploited to increase the floating point performance
of each node, at fixed remote bandwidth.
\begin{table}[hbt]\centering
\begin{tabular}{|c|c|}
\hline\hline
Linear lattice size & $\rho$ \\
\hline
$3^3$ & $7.5$\\
$4^3$ & $10$\\
$6^3$ & $15$\\
$8^3$ & $20$\\
\hline
$3 \times 6 \times 6$ & 11.25\\
$3 \times 4 \times 4$ & 9 \\
\hline\hline
\end{tabular}
\vspace*{5mm}
\caption{Local vs remote memory accesses: this table is the same
as the previous one, except that $\rho$ is estimated taking into
account the trick, described in the text, that reduces
remote accesses. The last two entries refer to non-square sub-lattices
that might be used when simulating a lattice of spatial size $48^3$
on large machines with $16 \times 8 \times 8$ or $16 \times 12
\times 12$ nodes.}
\label{tab:table_remote2}
\end{table}
Clearly the actual values of $\rho$ which can be accepted
must be studied more carefully (possibly simulating architectural
details of the mechanisms that hide remote communications)

\item
To discuss memory-size requirements in more details,
one has to distinguish between the case of full QCD simulations
and calculations in the quenched approximation.

In full QCD simulations, by far the largest amount of time is spent in the {\em
updating} process. In this case, on-line memory has to be large enough  to
allow for the implementation of efficient algorithms. State-of-the-art update
algorithms need a large number of auxiliary fields on each lattice site. We use
as unity the amount of memory associated to one fermion field (24 data words,
corresponding to 192 Bytes in double precision. We call this quantity a fermion
equivalent - feq - in the following). A generous estimate, leaving space for
more sophisticated, presumably more memory intensive algorithms, is about
$\simeq 200$ feq per site.

On the other hand, in the case of the quenched approximation, the updating
process may be neglected for both computing power and memory requirements (less
than $10$ feq per lattice site are needed). Instead, we have to consider the
memory requirement originating from the {\em measurement} of a heavy-light
form-factor. The database needed for such a calculation consists of one gauge
field configuration, one Pauli term, $N_l+N_h$ fermion propagators ($N_h$ and
$N_l$ are the numbers of heavy and light fermions respectively), each
replicated for the number of momenta and operator insertions used and for each
lattice site (typical cases, being $N_h = N_l = 4$, 3 momenta and one operator
insertion). Quenched QCD will be used essentially for heavy quark
phenomenology. Here the real problem is the extrapolation to the $b$ quark
mass. To be safe one should have a physical cutoff much larger than the masses
that enter the simulation. Then {\em large} lattices, of the order of $100^4$,
are necessary.

\begin{table}[bt]\centering
\setlength{\tabcolsep}{.2pc}
\begin{tabular}{|c||c|c|r|r|}
\hline
\hline
$ U_{ab}(x,\mu) $ & gauge fields & 72 W & 3 feq\\
$ S_{ab}^{\alpha\beta}(x,0) $ & fermion propagator & 288 W & 12 feq\\
$ \psi_{a}^{\alpha}(x) $ & (pseudo-) fermion field & 24 W  & 1 feq\\ 
$ (\sigma \cdot F)_{ab}^{\alpha\beta}(x) $ &
 Pauli term for improvement & 72 W & 3 feq\\

\hline\hline
\end{tabular}
\vspace*{5mm}
\caption{Data structures used in LQCD and corresponding memory
  requirements (in words and fermion equivalent storage) per lattice point.
  Greek indices
  run from $1$ to $4$ and Latin indices from 
  $1$ to $3$. The first three entries are general complex
  matrices, while the Pauli term is hermitian: 
  $(\sigma \cdot F)_{ab}^{\alpha\beta} = [(\sigma \cdot
  F)_{ba}^{\beta\alpha}]^*$. }
\label{tab:data_struct}
\end{table}

We summarize our memory requirements in table \ref{tab:data_struct} (where the
size of the relevant data structures are presented) and in table
\ref{tab:figures}, where actual memory sizes are collected, under the assumptions
of using double  precision\footnote{The necessity of double precision arithmetic
in full  QCD has been investigated in the literature \cite{double} and will not be
discussed here.} throughout. From the first two lines of table \ref{tab:figures},
we see that we cannot expect to keep the whole data-base in physical memory when
{\em large} lattices are considered. However, if only two propagators at the time
are kept in memory, for ease of programming, while the others are either
recalculated (the heavy ones) or stored and reloaded from disk (the light ones),
memory requirements reduce sharply (third line in the table).

We conclude that, by judiciously swapping data to disks, a memory size of the
order of $\geq 1$ TByte is a good compromise for both our case studies.
Alternatively, one might consider two memory options: a {\em small memory}
machine ($\simeq 500$ GByte) for full QCD and a {\em large memory} version ($1
\div 2$ TByte) for quenched studies.

\begin{table}[hbt]\centering
\begin{tabular}{|c|c|r|}
\hline\hline
Case & updating & measurement \\
\hline
{\em small} lattice, full QCD & $400$ G & $1.4$ T \\
{\em  large} lattice, quenched QCD & $200$ G& $13$ T \\
{\em large} lattice + disk & $200$ G & $1.8$ T \\
\hline\hline
\end{tabular}
\vspace*{5mm}
\caption{Total memory requirements for the case studies discussed in the text.
The line labeled {\em + disk} refers to the case in which two propagators only
are kept in memory (all others being swapped onto disk or recomputed).}
\label{tab:figures}
\end{table}

\item 
Fast input-output is mandatory, as obvious from the previous point, for studies
on large lattices. As a rule of thumb, we may want to load or store
one ({\em large} lattice)
propagator ($\simeq$ 250 GBytes) in little more than one minute.
This requires a global bandwidth of the
order of $2-3$ GBytes/sec.

For full QCD permanent storage of the configurations is required due to 
the computing effort needed to generate them. 
This is a storage-density (as opposed to bandwidth) problem
which is independent of the machine architecture and should be discussed
in a different context, with potential links with the GRID \cite{grid} project, likely
to be supported by the European Commission.
In the case of large lattices in quenched QCD the strategy of
computing {\em on the fly} without saving configurations is the best.
Only the final correlation functions are saved and this means at most a few
tens of MBytes per configuration.
\end{enumerate}

Processing performance is strictly speaking not a clear-cut requirement: the
more is available, the better. We can estimate how much is enough, however, by
extrapolating the present state of the art. A sustained performance of 300
GFlops (with perhaps 40\% efficiency) is now heavily used for full QCD
simulations on lattices of size $24^3 \times 48$ \cite{CPPACS}. If we assume a
critical slowing down where computer time grows like $a^{-7}$
\cite{CPUscaling}, we would like to have a sustained performance two  orders of
magnitude higher if we want to halve $a$. 

An ambitious target for our project is therefore a {\em total installed
performance} in the order of $10 \div 30 $ TFlops. From the point of view
of physics requirements, it is not important that this computing power be 
sustained on a {\em single} system. Several smaller machines can perform
equally well (or perhaps better), as long as {\em each of them} is able to
handle large enough lattices.

Also, we must envisage the operation of some lower performance (and
correspondingly smaller memory) machines, where small lattices are handled and
algorithms, programs and physical parameters are tuned before a large
calculation is moved onto a large production machine.

\section{The APEmille Project}
In this section we briefly review APEmille.

APEmille is the present generation  APE project.  It is based on the standard
structure of a large array of processing nodes arranged at the edges of a three
dimensional mesh and operating in SIMD mode. 

At present (November 2000),
several medium-size installations are up and running, while several
larger units are under construction (see table \ref{tab:statusape1000}).
Considering all large and small machines, the integrated peak performance
available at the end of the year 2000 will be about 1 TFlops at INFN
and about 400 Gflops at
DESY. The largest single system will have a peak performance of 250 (possibly
500 Gflops). Other institutions in Europe are procuring (or considering to
procure) APEmille machines.

\begin{table}[hbt]\centering
\begin{tabular}{|l|c|r|}
\hline\hline
Site & peak performance & status \\ \hline
Rome & 260 Gflops & running \\
Zeuthen & 130 Gflops & running \\
Rome II &  65 Gflops & running \\
Bielefeld & 80 Gflops & running \\ \hline
Milano/Parma & 130 GFlops & planned, Dec. 2000\\
Pisa & 130 Gflops & planned, Dec. 2000\\
Rome II & + 65 Gflops & planned, Dec. 2000\\
Rome & + 260 Gflops & planned, Dec. 2000\\
Zeuthen & + 260 Gflops & planned, Spring 2001\\
\hline \hline
\end{tabular}
\vspace*{5mm}
\caption{
A short list of some large APEmille existing installations and of plans
for the near future.}
\label{tab:statusape1000}
\end{table}

In a typical critical LQCD kernel (a solver for the Dirac operator) coded in the
high level TAO programming language,  measured sustained performance in single
precision is about 44\%  of peak performance and in double precision it is about
19\% of peak  single precision (i.e. 80\% of peak performance in double
precision). Higher efficiency can be obtained with more careful programming: we
have pushed single precision  performances up to about 58\% of peak speed
writing the key portions of the Dirac solver in assembly.

In a later section, we will claim that an architecture \`a la APEmille continues
to be a very good choice for LQCD computing. We see however a number of problems
in APEmille, all pointing to the development of a new generation system:
\begin{itemize}
\item{{\bf Peak Performance}}
APEmille machines can be made larger than we plan to build, but not too
large. The largest system that can be assembled with the present
hardware building blocks is a configuration of $8 \times 8 \times 32$ nodes,
corresponding to 1 TFlops peak performance (APEmille systems
can be configured in principle as $8 \times 8 \times 2n$ arrays).
Still larger systems
would need some minor hardware development and would probably be not
convenient in LQCD, since they have an unordinately large number of nodes
along one dimension.
\item{{\bf Memory Size}} The very large APEmille machine described above has 
64 GBytes memory. This is still several times lower than discussed in the
section on requirements.
\item{{\bf Floating-point precision}} APEmille is basically a single precision
machine (performance decreases by  factors from 2 to four in double precision).
However, in future large LQCD simulations double precision will be
necessary in an increasing number of cases.
\item{{\bf Little space for improvements}} APEmille is architecturally very
simple, since it relies on accurate and rigid hardware synchronization. This
style of synchronization is difficult to support if the system clock is
increased significantly. For this reason, we see little space for incremental
improvements in performance.
\end{itemize}

\section{A Review of Similar Projects}
In this section, we gather some information on similar projects, carried out by
other groups.  To the best of our knowledge, the following activities are in
progress:

\begin{itemize}
\item{{\bf CP-PACS}}

The CP-PACS collaboration have made a feasibility study of a future project
which follows the CP-PACS project. Extrapolating the data of the performance
obtained in recent full QCD simulations  on the CP-PACS computer,  they have
estimated the computer time required for a large-scale  full QCD calculation,
with the quality of data comparable to that of the present quenched QCD study
on the CP-PACS. They assume that lattice action and the simulation algorithm
are  identical to the present simulation on the CP-PACS. Their estimate, 
$\simeq$ 100 TFlops $\cdot$year \cite{iwasaki}, is somewhat larger than the one
of the ECFA panel.
In addition to their feasibility study, CP-PACS are carrying out basic
research on the following two topics, that they consider as very important
technologies for the next-generation of massively parallel computers:
\begin{itemize}
\item
Development of an architecture of high-performance memory-inte\-grated processor
for the next generation massively parallel computers.
\item
Establishment of a model of parallel I/O, parallel visualization and
man-machine interface, which can process efficiently and flexibly the enormous
amount of data generated by massively parallel computers.
\end{itemize}

CP-PACS hope they will have a chance to develop a next-generation computer
using the results of their  basic research in the near future, but they do not
have a project at the present time.

\item{{\bf Columbia}}
The Columbia group have officially embarked on the design and construction of
their next machine \cite{Norman}.  The design effort is still on a fairly high
level with choice of processor and communications technology being the first
questions that have been resolved.  Most significant is the choice of
microprocessor, which is provided by an IBM PowerPC core. This follows from an
arrangement with IBM that permits to exploit proprietary technology to
construct a full processing node  (memory included) on a single chip.  This
feature  provides the name to the new project QCD on a Chip (QCDOC).  
The node will contain a PowerPC 440 core, one 64-bit, 1 Gflops FPU (an
integrated part of the PowerPC architecture), 4 MBytes of embedded DRAM and 8
bi-directional serial inter-processor links, each operating at 0.5
Gigabits/sec.  If they are able to achieve this frequency, this would give a
total off-node communications bandwidth of 1 GBytes/sec.

The group is now busy  to determine the other details of the project and begin
the detailed design of the node.
\end{itemize}

We also include an arbitrary selection of two (out of the many) interesting
examples of PC-based cluster architectures for comparison.

\begin{itemize}
\item{{\bf The Wuppertal Cluster ALiCE}}

The "Institut f\"ur Angewandte Informatik" at Wuppertal
University has installed the first half of the
Alpha-Linux-Cluster-Engine (ALiCE) in 1999. When the system is fully
installed, in May 2000, it
will consist of 128 DS10 uni-processor workstations connected by
a Myrinet multi-stage crossbar switch. All CPUs will be upgraded to
600 MHz Alpha 21264 EV67 chips with 2 MBytes second level off-chip cache
\cite{wupper}.

The cluster is intended to perform efficiently in several HPC application
profiles at the University of Wuppertal, including computational chemistry,
electrical engineering, scientific computing and simulations of quantum field
theories. 

Of particular interest is the operability of this self-made system in  a
University's multi-user environment. In computer lab courses, the emphasis is
on "Physics by High Performance Computers". Several student groups use the
system simultaneously in interactive mode much alike a desk-top system.  

A forward looking ALiCE-project, to be carried out together with the developers
of the ParaStation communication software from Karlsruhe university, deals with
optimization of efficiency and data organization for ALiCE under real life
conditions, in particular with the goal to make parallel I/O and file system
functionalities available.

\item{{\bf PMS, The Poor Man's Supercomputer}}
A PC cluster has also been developed at E\"otv\"os University in Budapest
\cite{PMS}. The current version of PMS has 32 PC's. Contrary to the previous
example, the PMS project has developed QCD-optimized communication hardware.
They use dedicated add-on boards to establish physical communications between
nearest neighbour PC's in a three dimensional array. The actual configuration
of 32 PC's can be imagined as a $2 \times 4 \times 4$ mesh of processors. The
system uses a standard Linux operating system and the favoured programming
style is the well tested SIMD paradigm. 

The present version of PMS is shaped by the requirement to reduce costs
as much as possible. Indeed, PMS uses cheap AMD K6-2 processors (delivering
only 225 MFlops each) while the special purpose communication interface has a
bandwidth of just 2 MBytes/sec.  We consider the PMS as a very good trade-off
between the advantages offered by the use of general purpose systems and the
performance boost that dedicated hardware is able to provide.

\item{{\bf The MIT-Jefferson Lab project}} 
This project is organized in two phases. Phase I has
been submitted to DOE in march 1999 \cite{negele} for a 256-processor cluster
at JLab and a 64-processor cluster at MIT. The building block is a 4-processor
Compaq SMP node with a $750$ MHz Alpha 21264 chip, $4$ MB cache/processor and
$1$ Gbyte memory/SMP. The communication network is based on Myrinet switches.
They plan to achieve a cost per sustained MFlops lower than \$$10$.
The collaboration currently operates a 12 node prototype cluster equipped with
667 MHz Alpha 21264 at MIT. They plan to complete phase I by 2001. The
collaboration has a quite intensive physics program mainly devoted to nucleon
structure. Phase II of the project, not approved yet, foresees, in the years
2002-2005, multi-Teraflop machines from a combination of QCDOC and clusters.
They are collaborating with Compaq to explore resources available at that time
(EV7 and EV8).
\end{itemize}

\section{Technological Scenarios}
In this section we discuss forecasts about the state of the art for
several enabling technologies in
the years 2001-2002. We cover the following points:

\begin{enumerate}
\item basic digital VLSI technology.
\item memory technology.
\item data-links.
\item Off-the-shelf processors.
\item The Crusoe architecture.
\end{enumerate} 

We conclude the section with a discussion of the architectural implications of
the technology-driven choice of overcoming the strictly synchronous operation
of APE100 and APEmille. 

\subsection{VLSI Technology}

APEmille is based on a chip-set designed with a $0.5 \mu$ digital CMOS
technology. A second source for the chip-set has been established, using a more
advanced $0.35 \mu$ technology. In the next few years, $0.25 \mu$ and $0.18
\mu$ CMOS technologies will be readily available.

A comparison of some key features of the silicon technologies used in APEmille
and of a representative of both $0.25 \mu$ and $0.18 \mu$ technologies is made
in table \ref{tab:tech}.

\begin{table}[hbt]\centering
\begin{tabular}{|l|c|c|c|c|}
\hline\hline
Feature      & ES2 $0.5 \mu$ & Alcatel $0.35 \mu$ & UMC $0.25 \mu$ 
             & UMC $0.18 \mu$ \\ \hline
$V_{DD      }$       & 3.3 V         & 3.3 V & 2.5 V & 1.8 V \\
Gate delay   & 180 ps        & 100 ps& 75 ps & 36 ps \\
Gate density & $10$ K/mm$^2$ & $20$ K/mm$^2$ & $45$ K/mm$^2$  & $90$ K/mm$^2$ \\
Memory (1P)  & $11$ Kb/mm$^2$ & $25$ Kb/mm$^2$ & $44$ Kb/mm$^2$ & $85$ Kb/mm$^2$\\
Memory (2P)  & $6$ Kb/mm$^2$ & $8$ Kb/mm$^2$ & $16$ Kb/mm$^2$ & $30$ Kb/mm$^2$ \\
Power/gate   & $0.5 \mu $W/MHz        & $0.4 \mu$W/MHz  & $0.2 \mu$W/MHz 
             & $0.1 \mu $W/Mhz \\
\hline\hline
\end{tabular}
\vspace*{5mm}
\caption{A summary of some key parameters for digital silicon technologies
used in APEmille and proposed for {\bf apeNEXT}. All values are directly obtained
from the relevant silicon foundries, except for the bit density of 1 Port or
2 Ports memory arrays in the UMC technology.
The latter are based on conservatively applied scaling rules}
\label{tab:tech}
\end{table}

The figures quoted in the table refer to processes that are (or will be)
readily available through the same European silicon broker that helped us
develop the second source of the APEmille chip set. 

Let us consider a scaled version of the APEmille processor. If we use a $0.18
\mu$ process, it should be easy to reach a clock speed between three to five
times higher than in APEmille, while we may expect to squeeze up to 9 times
more transistors onto the same silicon area. We can stay on the safe side
planning to use a clock  frequency of $200$ MHz. An LQCD optimized processor
running at this clock frequency with one floating-point pipeline would peak at
1.6 Gflops , using  the well known {\em normal} operation $ a \times b + c$,
performed on complex operands. A chip three times more complex than J1000 (and
three times faster) would dissipate less than two times more power.

\subsection{Memory Technology}
We limit ourselves to memory systems used in future high-end PC's or low-end
workstations. This choice (the same as APE100 and APEmille) should be the most
effective to provide the highest level of integration, reduce costs and
guarantee part availability.

In the near future, planned memory systems are either  RAMBUS DRAM's
or DDR SDRAM's.\footnote{In the following we do not distinguish 
between traditional DDR SDRAM and new ``flavour'' DDR SDRAM like Sync-Link 
because both are evolutionary designs of the same basic structure.}

The DDR SDRAM (Double Data Rate Synchronous DRAM), is the evolution of  the
mature SDRAM (Synchronous DRAM) technology (widely used  in the APEmille
machine). The SDRAM is a low latency burst oriented device  made of multiple (2
to 4) banks of asynchronous DRAM controlled  by a synchronous controller which
allows pipelining of the  I/O interface (one word is accessed for every clock
cycle).  The Double Data Rate architecture realizes two data transfers per
clock cycle using both edges of the clock and one special reference  signal to
fetch corresponding data.

The RAMBUS is a more advance memory architecture which works as a  chip-to-chip
system-level interface rather than a conventional memory device.  The RAMBUS
RDRAM (which stands for RAMBUS Direct Dram) shares the same  architectural idea
of the SDRAM one,  a core asynchronous plus a synchronous controller.  It makes
use of a large degree of parallelism (32 interleaved memory banks)  on a narrow
internal bus.  The RAMBUS RDRAM is based on the Direct RAMBUS  Channel, a high
speed 16-bit bus at a clock rate of 400 MHz, which thanks to  the adoption of a
dedicated signalling technology (RAMBUS Signalling Level) allows 600 MHz to 800
MHz data transfers.  

In table \ref{tab:memory_compare} we summarize the main features of the two
technologies, for currently available and next generation (less than 2 years
from now) chips.

\begin{table}[hbt]\centering
\begin{tabular}{|c|c|c||c|c|}
\hline\hline
 & DDR & RDRAM & DDR & RDRAM \\
\hline
Data rate              & 200 MHz  & 800 MHz & 400 MHz & 800 MHz \\
Memory size  & 256 Mbit & 128/144 Mbit & 1 Gbit & 256 Mbit \\
Organization & x4,x8,x16 & x16,x18 & x16,x32  & x16,x18 \\
Peak bandwidth & 0.4 GB/s (x16) & 1.6 GB/s & 1.6 GB/s (x32) & 1.6 GB/s \\
Package & TSOP(66) & BGA & TSOP(80) & BGA \\
Power (VCC) & 2.5 V & 2.5 V & 1.8/2.5 V & 1.8 V \\
I/O type & SSTL2 & RSL & SSTL (?) & RSL (?) \\
Power cons. & 80 mA & 330 mA & ? & ? \\
Cost (norm.) & 1.0 & 1.8 & ? & ? \\
Sample/Prod. & Now/Now & Now/Now & 3Q99/4Q00 & ? \\
\hline\hline
\end{tabular}
\vspace*{5mm}
\caption{A summary of several important figures for two options of dynamic
RAM's. The second and third columns refer to presently available DDR and RAMBUS
devices. The fourth and fifth columns refer to the expected evolution of these
devices in the next two years.}
\label{tab:memory_compare}
\end{table}

Some comments are in order:
\begin{itemize}
\item The simple architecture of the DDR SDRAM allows  larger memory size per
device. For a given fixed amount of memory, this reduces  the number of used
components.
\item Since power  consumption is proportional to the interface clock  (a
factor 4 between RAMBUS e DDR), aggregated memory systems using  the DDR SDRAM
reduce the global consumption. 
\item On the other hand the extremely high peak bandwidth of the RAMBUS allows
to build a very  fast memory system with minimum impact on board space
occupancy  (compact BGA packaging).
\item The logic complexity of a RAMBUS interface is much larger than for a
DDRAM controller (the latter could be easily designed on the basis  of the
experience done in the realization of the APEmille memory controller). On the
other hand, several silicon foundries make a RAMBUS controller available as a
core cell.
\end{itemize}

We conclude this section by presenting in table \ref{tab:memory_systems} two
possible DDRAM-based memory systems for {\bf apeNEXT}. The performance target
is set by our basic performance figure, discussed in the previous subsection of
$1.6 Gflops$ and  $R = 4$, leading to a bandwidth requirements of at least $3.2$
GBytes/sec (assuming double precision data words throughout).

\begin{table}[hbt]\centering
\begin{tabular}{|l|c|c|}
\hline\hline
chip-size          & 1 Gbit         & 1 Gbit\\
chip  organization & 32 bits        & 32 bits \\
chip number        & 4              & 2\\
word size          & 128 bit        & 64 bit\\
bank size          & 512 MBytes     & 256 MBytes\\
frequency          & 300 MHz        & 400 MHz\\
total bandwidth    & 4.8 GBytes/sec & 3.2 GBytes/sec\\
power consumption  & 640 mW         & 400 mW\\
\hline\hline
\end{tabular}
\vspace*{5mm}
\caption{Basic features of two possible memory systems for {\bf apeNEXT} based
on DDRAM memory technology. Power  consumption is estimated by rescaling
data available for present generation systems}
\label{tab:memory_systems}
\end{table}

In conclusion, forthcoming memory technology is adequate to support the
processor performance discussed above. There is in fact reasonable space to
consider either fatter node processors, or multi-processor chips.

\subsection{Data-link Technology}

We now consider {\em remote communications} which, in our opinion, is a key
technological challenge for the project. 

Assuming our reference figures - 1.6 Gflops per node, along with $R = 4$, and
$\rho = 8$ (as defined in the previous sections) - we require an
inter-processor communication  bandwidth of about 400 MBytes/sec. As discussed
above, several code optimization steps are able to reduce the amount of data to
be transferred. The overlap between computation and communication can also be
increased. All this steps reduce bandwidth requirements. We will stick however
to the previous figure, so a large safety margin is established.

The needed communication patterns are however very simple: communications are
needed between nearest-neighbours  (L-shaped paths, between next-to-nearest
neighbours are also useful) in a 3-d array of processors, where each processor
has 6 direct links to its nearest neighbours. The real challenge in this area
is therefore more the implementation of a fast, reliable and cheap link than
the development of any clever routing strategy.

In APE100 and APEmille, links use large, parallel and synchronous data paths.
Data words are injected at the transmitting end of the line following a rising
transition of the clock and are strobed into the receiving end of the line
at the next rising edge of the clock. This works if
\begin{eqnarray}
T_t < T_{clock} \\
\Delta T_{clock} << T_{clock}
\end{eqnarray}
where $T_t$ is the travel time over the physical link, $T_{clock}$ is the clock
period and $\Delta T_{clock}$ is the phase spread between (nominally
aligned) clock signals at various places in the machine. The conditions are met
in APEmille, where $T_{clock} = 30 ns$, $T_t \simeq 10 ns$ and $\Delta
T_{clock} \simeq 4 ns$, while they become clearly unrealistic for frequencies
of $ \simeq 200$ MHz.

More advanced (high bandwidth) link technologies have recently become available,
in which data and timing information are both encoded on the physical link, so
asynchronous operation is possible. In the bandwidth range relevant for us,
we have considered three different options:
\begin{itemize}
\item{{\bf Myrinet-like links.}} The physical layer of the Myrinet interconnect
uses low swing single-ended signalling. One byte is encoded onto ten signal
lines, carrying also timing information. The full duplex link uses two such
busses. The present generation Myrinet link has a bandwidth of 160 MBytes/sec
(using both edges of an 80 MHz clock),  while a new generation (Myrinet-2000,
320 MBytes/sec) is under test. The main advantage of Myrinet links is that they
pack a lot of bandwidth while keeping operating frequency low. Board layout
details, connectors and cables are also very well tested. We are informally
discussing with Myricom the possibility to use this link for {\bf apeNEXT}.
Myricom have agreed to allow us to use the link level (SAN-port) circuitry for
their latest Myrinet chips (Myrinet 2000) as a basis for the {\bf apeNEXT}
links. Under a suitable non-redistribution agreement, Myricom will make
available to the collaboration the layout of the basic cells, along with their
Verilog models.
\item{{\bf LVDS based links.}} The Low Voltage Differential Signalling (LVDS)
technology is now widely used in many telecom and network technologies, like
the Scalable Coherent Interface (SCI). LVDS is designed to work up to $\simeq
622$ MHz. Several redundant encoding schemes (e.g., 8 bits into 10 bits) have
been proposed. LVDS cells are readily available from several silicon vendors.
New generations FPGA chips have been announced including LVDS options. Work is
in progress to test LVDS links, as described later in this document.
\item{{\bf High speed proprietary links.}} Several silicon houses (e.g., Texas
Instruments (TI), National Semiconductor(NSC), LSI Logic)  have developed very
high speed proprietary links, aimed at the Gb Ethernet, Fiber-Channel,
Infini-Band markets. The typical bandwidth is higher than 1 Gbit/s. Complete
encoding-decoding black-boxes are usually available. This option has two main
drawbacks: it makes the whole project dependent on a specific silicon house,
and requires very careful layout of the printed circuits and proper choices of
cables, connectors and the like.
\end{itemize}

Basic figures of the three options are compared in table \ref{tab:links}, where
we use for the LVDS case a more conservative frequency of 400 MHz.

\begin{table}[hbt]\centering
\begin{tabular}{|l|c|c|c|c|}
\hline\hline
Technology        & Frequency & Pins & Bandwidth & Power Dissipation\\ \hline
Myrinet           & 160 MHz   & 20   & 320 MBytes/s & 300 mW \\
LVDS              & 400 MHz   & 40   & 400 MBytes/s & 200 mW \\
SerDes(TI)        & 1.24 GHz  & 10   & 400 MBytes/s & 400 mW \\
NSC DS90CR483/484 & 784 MHz   & 18   & 672 MBytes/s & 1500 mW \\
\hline\hline
\end{tabular}
\vspace*{5mm}
\caption{Basic figures for several link technologies. All figures refer to full
duplex links. An 8 bit into 10 bit encoding has been assumed for the LVDS
case.}
\label{tab:links}
\end{table}

An important issue is the reliability of the network, usually measured in BERR
(average number of errors for transmitted bit). If we require fault-less
operation of a large machine for one day (say, 2000 links active for 50\% of
the time), we need a very low value of $BERR \simeq 10^{-17}$. For
comparison's sake, {\bf measured} stable operation of an APEmille machine with
250 nodes for periods of a few days implies $BERR \le 10^{-15}$.

Machine reliability greatly improves if the network is able to recover from
network errors by retrying a failed communication (this impacts on link
latency, but the impact can be made low with some care). For instance  a 
comfortable $BERR \simeq 10^{-12}$ implies that one communication must be
retried on the machine every second.

The above discussed feature however requires some degree of non asynchronous
operations, with important technological implications. Regardless of the
technological choice made for the processor, we think that no real advantage is
gained by departing from the Single Instruction Multiple Data (SIMD) or Single
Program Multiple Data (SPMD) programming style used in previous generation APE
machines. At the hardware level, APE processors of all previous generations
have been hardware-synchronized with an accuracy of a fraction of clock cycle.
Although logically very neat, this is  rapidly becoming impossible, for clock
frequencies  higher than 100 MHz and across physical scales of several meters.
We consider an approach  in which independent processors, while running at the
same frequency, are only loosely synchronized. Logical synchronization will
have to be enforced by some form of software-controlled barrier.

\subsection{Off-the-Shelf Processors}  
In this section, we briefly consider of-the-shelf processors as a potential
building block for the computational core of {\bf apeNEXT}. With one notable exception
(see later), we choose to consider only the option of using commercially
available boards (in other word, if we decide to use a commercially available
option, we want to drop altogether any hardware development not involving the
network). In the following section we will compare the relative merits of
off-the-shelf versus custom processors.

Standard off-the-shelf processors have increased in performance by more than
one order of magnitude in the last 8-10 years, with an even more remarkable
improvement in the efficiency of floating point computations. 
Standard PC boards using off-the-shelf processors have been used for small
scale LQCD simulations. The relevant codes are written in familiar programming
languages, like C (or C++) or Fortran. Efficiencies are limited by
bottlenecks in memory access as soon as the data base involved in
the computation exceeds the cache size (which is the typical situation
in realistic LQCD simulations). Measured efficiencies on a Pentium II processor
running at 450 MHz are of the order of $30\%$, for real-life production
programs (running on just one node, i.e., with no communication overheads)
\cite{reiner_and_friends}.

A detailed discussion of the expected technical road-maps for 
off-the-shelf processors in the next few years in general terms
would exceed the scope of this document. Instead, we discuss 
the features of a typical high end microprocessor, that might be 
used today and apply usual scaling laws. For this purpose, we 
(rather arbitrarily) take the AMD Athlon. A number of features 
relevant for LQCD simulations are shown in table \ref{tab:amd}.

\begin{table}[hbt]\centering
\begin{tabular}{|l|c|}
\hline\hline
Clock frequency                 & 600 - 1000 MHz \\
FP ops (single precision)       & 4 per clock cycle\\
FP ops (double precision)       & 1.6 per clock cycle \\
FP latency                      & 15 clock cycles \\
L1 Data Cache                   & 64 kBytes\\
Data bandwidth to L2 cache      & 1.6 GBytes/sec \\
Sustained LQCD performance       & 360 MFlops \\
Power consumption (750 MHz)     & 35 W \\
Retail price (600 MHz)          & 200 Euro \\
Retail price (750 MHz)          & 375 Euro \\
Retail price (800 MHz)          & 500 Euro\\
\hline\hline
\end{tabular}
\vspace*{5mm}
\caption{Features of the AMD Athlon processor relevant for LQCD simulations.
Sustained performance is estimated under the assumptions discussed in the text.}
\label{tab:amd}
\end{table}

\begin{figure}[tbh]
\epsfig{file=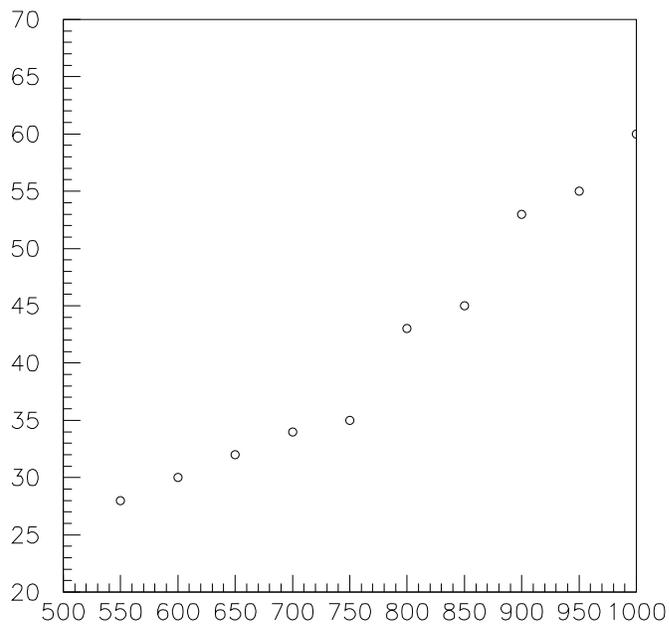, width = 4.2in}
\caption{Power consumption (W) of the AMD Athlon processor as a
function of the clock frequency (MHz) \cite{AMD}.}
\label{fig:AMD}
\end{figure}

In the rest of the discussion, we consider the version of the Athlon
running at 750 MHz. Indeed, figure \ref{fig:AMD} shows that for higher
frequency, power dissipation increases faster than linearly. 

If we assume an efficiency comparable to the one measured on Pentium systems,
we expect a sustained LQCD performance of $\simeq 360 Mflops$ per processor,
making it possible to use dual-processor 
mother-boards without jeopardizing efficiency
(a quad-processor system would saturate the maximal theoretical bandwidth 
of 1.6 GBytes/sec to access a memory bank working at 200 MHz  assuming our 
usual value of $R \simeq 4$).

In summary, a high end PC-like node should be able to sustain a performance
of $\simeq 700 Mflops$ running LQCD codes in double precision.
We can take this as our basic
building block, with just a few relevant figures summarized in 
table~\ref{tab:amd2}

\begin{table}[hbt]\centering
\begin{tabular}{|l|c|}
\hline\hline
Sustained performance  & 700 MFlops \\
Power dissipation      & 90 W \\
Tag price              & 1500 Euro \\ 
\hline\hline
\end{tabular}
\vspace*{5mm}
\caption{Basic figures for a PC-based node of an LQCD engine, using currently
available off-the-shelf hardware.
Price estimates are
made at current retail prices. They include 512 MBytes main memory.
No LQCD networking or infrastructure is considered.}
\label{tab:amd2}
\end{table}

This nodes needs a sustained interface to neighbour nodes in the three
directions of the lattice grid with a bandwidth of 
$\simeq 200$ MBytes/sec. 

In conclusion, a system delivering 1 TFlops sustained LQCD performance would cost
more than  2.2 MEuro in processors only, 
and dissipate more than 130 KW power. We will discuss the
implications of these numbers in the following section.

\subsection{The Crusoe Architecture}

\begin{figure}[tbh]
\epsfig{file=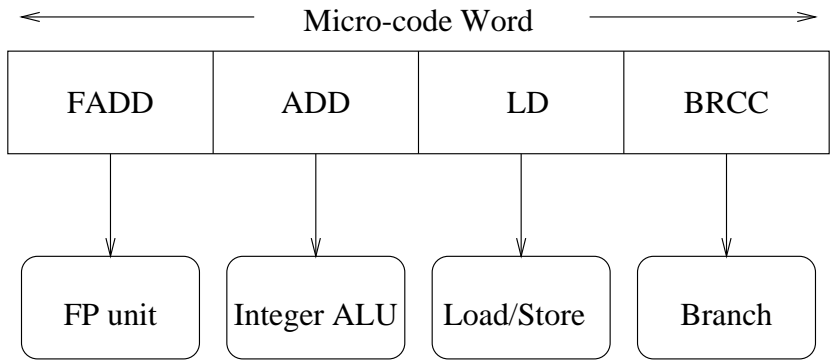}
\caption{The Crusoe architecture (adapted from \cite{crusoe}.}
\label{fig:crusoe}
\end{figure}

Very recently a new processor architecture (known as the Crusoe) has been
proposed by Transmeta Corporation. The Crusoe is advertised as as a streamlined
(hence very low consumption) processor, optimized for laptops or other mobile
computers. The Crusoe has a very simple architecture, that, when used
behind a core-level software environment, emulates
the Intel X86 architecture. From our point of view, it is more relevant
that the architecture of the Crusoe is extremely similar
to the combination of the
processing chips used in APEmille. Basically, the Crusoe core is a micro-coded
system in which several functional units operate concurrently on data coming
from a medium-size register file (see figure \ref{fig:crusoe}). The chip has also
a data instruction cache, as well as two different memory interfaces.

A high-end implementation of the Crusoe (advertised as available
from Summer 2000) is called the TM5400. It runs at 500 (maybe 700) MHz and
dissipates about 2.5 W, when running at full speed.

At present, no Crusoe-based boards are available. It is likely that the first
commercial products using Crusoe processors will be laptop machines, that
obviously do not meet our requirements. We have therefore to consider the
option of building a Crusoe-based {\bf apeNEXT} processing board.

The main advantages of this choice are basically summarized by saying that we
would be using an architecture very similar to APE, while being spared
the burden of designing our own processor.

We have contacted Transmeta to explore this option. They
stated that:
\begin{itemize}
\item They are not ready to provide critical
details of the internal architecture (for instance, no information was
provided on how many floating point operations can be executed at each
clock cycle).
\item Sufficient details of the VLIW core will not be given. Indeed
Trans-Meta attitude is that all programming for the Crusoe must be done at
the level of the 
Intel architecture, and must be translated with their proprietary
software.
\end{itemize}

With these pieces of information available and considering also that:
\begin{itemize}
\item
It is not clear whether chips can be procured at an early enough stage of
the project. 
\item
It is not obvious how fragile the whole Crusoe endeavour is.
\end{itemize}
we think that the present situation does not suggest
to base a new project on Crusoe. Of course, we will keep a close watch on any
related development. 

\section{Custom or Off-the-Shelf Processor} 
Previous generation LQCD projects have used either custom processors,
or substantial enhancements to standard processor architectures or
processors developed for niche applications. No big project has been based on
standard off-the-shelf processors so far.
Today, a decision to follow the same path is not as obvious as it has been
in the past, since off-the-shelf processors have increased in performance by
more than one order of magnitude in the last 8-10 years, with a remarkable
and even more relevant improvement in the efficiency of floating point 
computations. 

In table \ref{tab:chefare} we compare a few numbers relevant for APEmille,
for the PC-based solution discussed in the previous section and for a
custom-based {\bf apeNEXT} architecture (in this case, we use several tentative
numbers discussed in early sections).

\begin{table}[hbt]\centering
\begin{tabular}{|l|c|c|c|}
\hline\hline
---- & APEmille & {\bf apeNEXT}: PC-based & {\bf apeNEXT}: custom\\
\hline
Peak performance         & 500 MFlops   & 1200 MFlops  & 1600 MFlops\\
Sust. performance        & 250 MFlops   & 360 MFlops   &  800 MFlops\\
Power Dissipation        & 1.5 W        & 35 W         & 3.0 W\\
\hline\hline
\end{tabular}
\vspace*{5mm}
\caption{Comparison of several key figures for APEmille processors
and possible options for {\bf apeNEXT}. We assume that a next generation 
custom processor has the same efficiency as APEmille.}
\label{tab:chefare}
\end{table}

No clear cut best choice emerges from these numbers. In
general, we see advantages both in custom architectures and in PC-based
architectures:

We believe that a custom architecture is superior for {\em very
large ($\geq 500 nodes$) systems} for the following reasons:

\begin{itemize}
\item lower power consumption by one order of magnitude.
\item significantly more compact mechanical design.
\item better scalability once the basic units are operating (reliability 
      and software issues of large systems).
\item easier interfacing with the necessary custom remote communication 
      network and the host system.
\item better control of technological aspects and less dependence on 
      changing commercial trends during the realization of the project.
\end{itemize}

On the other hand, we see several advantages stemming from the use of
PC-derived systems for {\em smaller machines}:

\begin{itemize}
\item
limited hardware development effort.
\item
standard software is readily available for major parts of the
compiler and the operating system.
\item
short lead time to commission a prototype system.
\end{itemize}

We see at this point the need to make a clear decision between
the two options: we decide to focus on the development of a LQCD
architecture based on an APE-like custom processing nodes, whose
architecture is described in the next sections. We base our decision on
the following points:
\begin{itemize}
\item we want to focus our project onto machines with very large 
performance. As explained earlier on, we will have to put
      together
several machines to really arrive at a VERY LARGE scale.
\item we think to be able to rescale and reuse a large wealth of building
blocks from APEmille, reducing the design time.

\item We think that the commissioning of a very large PC-based system
(involving several thousand PC's all over the collaboration)
is a huge (and new for us) project in terms of hardware (thermal and power
management, availability in case of hardware failures)
and software (control of a large network) issues for which we
have no real background.
\end{itemize}

We obviously think that a PC-based system is still a viable alternative
(discussed at some length in the preliminary proposal) for small or
medium-scale systems.  At this point in time, we do not consider however the
development of such a PC-based cluster as a priority for the {\bf apeNEXT} project.
We are however willing to collaborate with any such project, making any
{\bf apeNEXT}-proper development that might be useful for a PC-based LQCD cluster
readily available for such purpose. To this end, two points are most important:
\begin{itemize}  
\item We plan to design the network processor, supporting LQCD-optimized
point-to-point communication in such a way that it can be easily interfaced to
a PC (say across a PCI interface). See the section on the network architecture
for more details on this point.  
\item We start from the beginning the development of a programming environment
that allows easy porting between PC-clusters and {\bf apeNEXT} systems.
\end{itemize}

\section{Architecture Outline} 
In this section, we outline an architecture, leading to stand-alone
{\bf apeNEXT} systems scalable from about 100 Gflops to about 6 TFlops peak
performance. 

Just one such high-end machine would offer a ten-fold increase in peak
performance with respect to currently available systems. Several (5 to 10)
high-end machines, working together with a comparatively larger number of
low-end systems, would allow to complete the physics program outlined in 
previous paragraphs. 

We propose the following structure:
\begin{itemize}
\item
a three dimensional array of processing nodes, linked together
by nearest-neighbour links. Each node is a complete
and independent processor.
All nodes execute the same program and are {\em loosely
synchronized}, i.e., they are started at approximately the same time
and proceed at approximately the same pace. They synchronize
when requested by the logical consistency of the program
(e.g., before exchanging data).

\item
Remote communications use FIFO-based weakly asynchronous connections
between neighbouring nodes. The SIMD/SPMD programming
style \`a la APE does not require complex handshaking protocols, since
transmitting nodes may assume that the receiving partner is always ready
to receive the incoming message.

This  simple mechanism brings 
several architectural advantages:
\begin{enumerate}
	\item It allows to use for the remote communications a programming
	style which is very similar to APE100/APEmille. The latter has 
	the very convenient feature that no explicit distinction between  local
	and remote memory accesses is required when coding a program.
	\item This programming style can be easily modified to allow hidden
	data transfers (data are moved on the links while the processing node
	is performing calculations).
	\item It drastically simplifies the global hardware synchronization
	logic of the  system.
\end{enumerate}

\item
The communication interface is in principle an independent component.
As discussed, the communication interface is conceptually based on
FIFOs, allowing "elastic" connections between nodes. This novel feature
has to be carefully simulated, but no serious problem is anticipated here.
We need a  fast, yet cheap and reliable\footnote{As already stated, due to
asynchronous operations of the machine, requirements on the bit-error rate of
the communication system are less demanding than in previous APE generations,
since it allows for repetition of transfers with minor performance loss.}
data-link. Using $\rho \simeq 8$, we need links of $400$ MByte/sec. As
discussed in the section on technology, two or three different solutions are
available.
\end{itemize}

As discussed in the previous section, we focus our project on an implementation
of the above outlined architecture based on a closely packed array of custom
processors. We have in mind an implementation allowing to build systems of
between about 1000 to about 4000 processors, along the following lines:

\begin{itemize} 
\item Each node is based on a VLSI processor running at  about 200 MHz clock.
The processor merges the functions of the control (T1000) and floating-point
(J1000) processors of APEmille on a single chip. Each node has a private memory
bank, based on commodity chips. Memory size per node is likely to be in the
range 256 MBytes - 1 GByte per node. The actual choice may be heavily affected
by cost factors. The basic floating point instruction is the {\em complex
normal} operation, so peak performance is 1.6 Gflops (double precision). As
already remarked, this requires a memory bandwidth of 3.2 GBytes/sec  ($R = 4$).
We are studying the possibility to increase performance by factors $2 \div
4$, by using some form of super-scalar or vector processing, in which several 
normal operations are performed concurrently.

\item A typical large system has between $8 \times 8 \times 16 = 1024$ and $16
\times 16 \times 16 = 4096$ nodes. We assemble nodes on processing boards,
similar to APEmille. Each processor is more compact than in previous
generations, and glue logic is almost completely absent.

\item The node (and the network) should support not only data transfers between
memory and register (as available on APE100 and APEmille), but also register to
register. This can be used to reduce bandwidth requests by splitting a complex
computation on more nodes, each node using local data as much as possible,
as remarked earlier.

\item A host system similar to the one used in APEmille is a possible choice
for the new machine. Based on networked Linux PC's and the CPCI bus, it is
mechanically compact and reliable.  Each PC will be in charge of several
boards. The actual number of boards connected to each PC is dictated by the
bandwidth available on the PCI bus to move data from APE to disk and
vice-versa. For the sake of definiteness, assume a system distributed on 
approximately 100
boards, with a total bandwidth of 2 GBytes/sec (that is 20 MBytes/sec
per board).
In this case, up to 4 boards can be handled by present generation CPCI
CPU's. Higher performance PCI busses (double size and/or double speed) may
allow to increase the number of boards connected to each PC. The host PC's will
be networked with the most appropriate technology available in due time.

\item We plan to take advantage of all handles offered by the non fully
synchronous structure of the machine to relax the requirements and to
simplify the structure of the host to {\bf apeNEXT} interface.

Basically we will hook the interface to just one or two nodes belonging to
each {\bf apeNEXT} board. (This can be done conveniently by connecting to
the corresponding network interface).
All complex patterns of input/output data movements, for instance
relevant to a write onto disk of a ``slice'' of {\bf apeNEXT} processors, are best
performed by assembling the data words onto the input/output nodes under
program control, and then issuing a single data transfer to disk.

We can load executable programs in a similar way, by first moving the code
to the input/output nodes and then having a ``loader program'' to move the 
data onto the whole array.

We need a lower level system able to access all nodes independently even
if the neighbouring nodes do not work correctly. This system is needed for
debugging and test purposes and (for instance) to start the ``loader''.
Speed is not  relevant in this case, so well tested standard systems
(such as the JTAG interface) would perfectly fit our needs.

\item We note that it is a relatively easy task to design the (fully
self-contained) processing node(s) in such a way that they can be connected to
a standard PCI desk-top PC. This possibility is very appealing for program
debugging and small scale application. We plan to pursue this design
characteristic.
\end{itemize}

In the following sections, we describe in more details some key components of
our new system.

\section{Architecture of the Custom Node}

In this section we present the architecture of a simple  custom node
for {\bf apeNEXT}. The main idea guiding our design has been that of reusing
bits and pieces of APEmille as long as possible, while making use
of technologic improvements to rescale performance as much as possible.
We chose these guidelines in order to shorten the design cycle.

The custom node on which a large scale {\bf apeNEXT} system is based is called
{\bf J\&T}, since it combines the functionalities provided in APEmille by the
control processor (T1000) and the mathematical processor (J1000). The  combined
processor shares just one memory bank.

\begin{figure}[tbh]
\epsfig{file=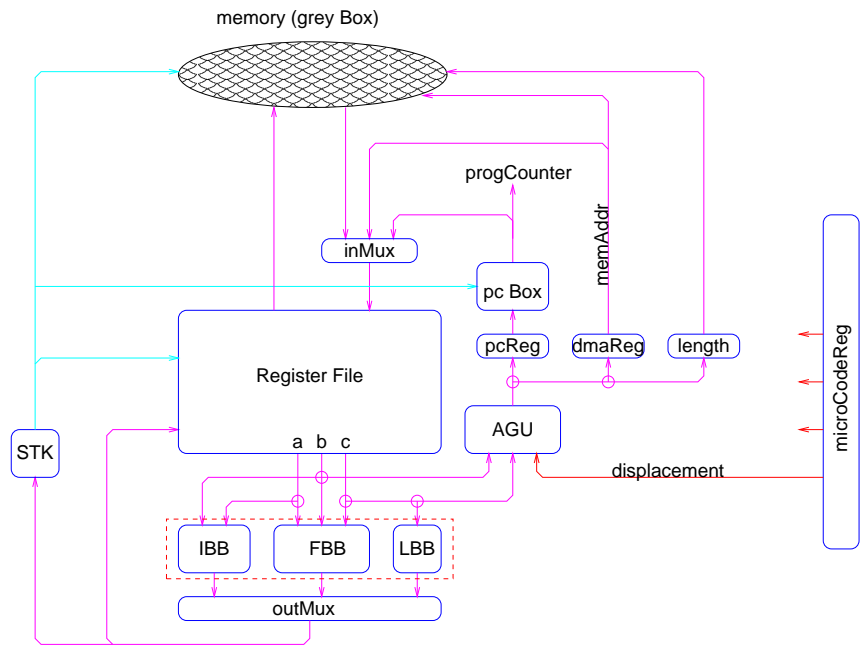}
\caption{Architectural block diagram of J\&T}
\label{fig:node}
\end{figure}

A basic block diagram of the
architecture is shown in figure \ref{fig:node}. The picture does not
cover in detail the memory and network interface. These points will be
discussed later on.
J\&T is centered around the register file, whose structure is
the same as the one used in APEmille. Data are transferred from memory to
register file (and back) through a bi-directional port.
Data available on the register file can be operated upon in just a few ways:
\begin{itemize}

\item Data words can be fed to the mathematical processor (the set of three
units within the red dashed frame). The latter contains a floating-point
data-path (Floating-point Building Block, FBB), an integer arithmetic unit 
(Integer Building Block, IBB) and a further unit providing first
approximations of some useful mathematical functions, such as $\sqrt{x}, 1/x,
\exp{x}$. This block is known as a Look-Up-Table Building Block
(LBB) in APE jargon. Results
of the mathematical block are written back to the register file  (for later
reuse or store onto the memory).
\item Data can be moved to the Address Generation Unit (AGU), where a memory
address or a branch-address can be computed out of two register-operands and
one immediate-operand (the displacement). New (data or branch) addresses are
stored in appropriate registers to be used at the next memory reference or
branch.
\item Logical tests can be evaluated on results computed from the mathematical
processor. The outcome of such tests goes onto a stack where more
complex logical conditions can be evaluated. The top of the stack is used to
control program flow by acting on the program-counter circuitry
(corresponding to {\tt if (...) then } in high level programs) or
to block write operations onto memory or register file ({\tt where (...)}
statements in APE-like high level programs).
\end{itemize}
The processor is controlled by a relatively large program word (called the
Micro-code Word) directly controlling the various devices in the node. (Almost)
no instruction decoding is performed on chip. This scheme has been successfully
used in the node processors of both APE100 and APEmille. A word size of
128 bits is large enough to control the system.

In the following, we describe in more details several key units of the
processor.
\subsection{The memory Interface and the Network Interface}
\begin{figure}[tbh]
\epsfig{file=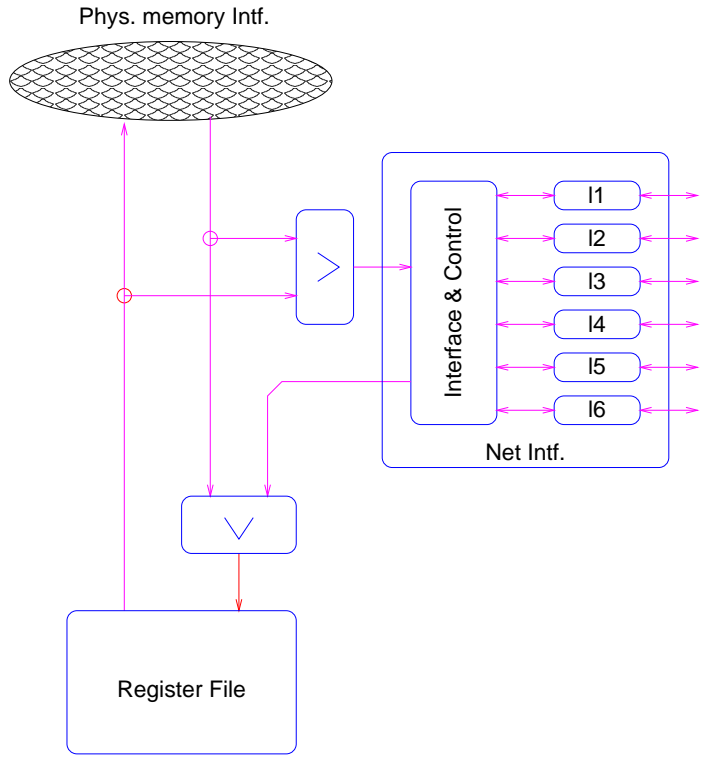}
\caption{Top level block diagram of the memory and network interface.}
\label{fig:memory1}
\end{figure}

In this section we describe the memory and network interface, sketched
as memory grey box in figure \ref{fig:node}. A basic structure of this
subsystem is shown in fig \ref{fig:memory1}.
The diagram shows several paths:
\begin{itemize}
\item there is a direct data path from the Register File to
the physical memory interface (and vice versa), supporting normal
memory access.
\item Data from memory can be also fed to the Network Interface
(and eventually routed to a remote node). Conversely, data arriving
from the Network (from a remote node) can be routed to the Register File.
\item Data words may be sent to the network from the register file. This is
a novel feature, allowing register-to-register remote communications.
This feature reduces remote bandwidth requests in some cases (notably
in the evaluation of the Dirac operator).
\item The network interface receives data from the memory (or the registers)
and route it to the appropriate destination through one of the six links
(Details on the network itself will be provided later on).
\end{itemize}

\begin{figure}[tbh]
\epsfig{file=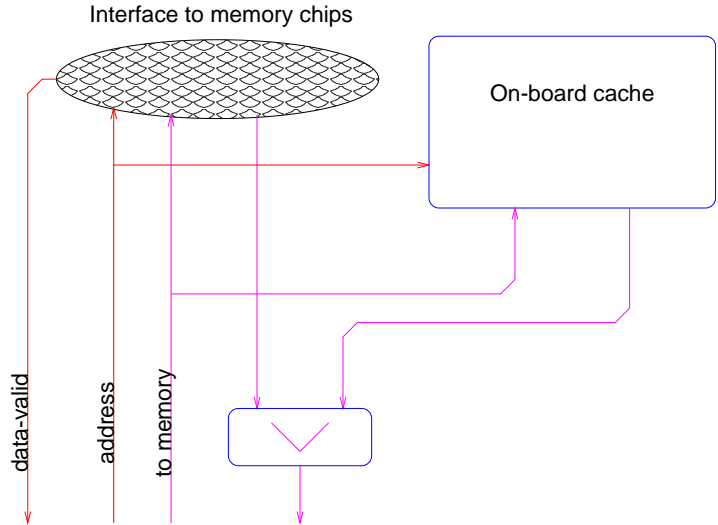}
\caption{A lower level view of the memory system.}
\label{fig:memory2}
\end{figure}

In figure \ref{fig:memory2} we show a more detailed view of the interface to
the physical memory.
We see that memory is divided into cache memory and external memory:
\begin{itemize}
\item {{\bf external memory}}. External memory implements the large memory
bank of the node. As discussed in the section on technologies, we may use
(for instance) DDR 1 Gbit memory chips. We have several options of
memory bus width and bank size satisfying bandwidth constraints
(see table \ref{tab:memory_compare}).
We want to leave these options open at this point in time.
For this reason, figure \ref{fig:memory2} still has a grey box. This
grey box contains the actual state machine controlling memory access,
memory correction circuitry, refresh control circuitry and any other
ancillary logic. The box will be designed in detail at a later stage of the
project, after the actual memory technology has been selected. For the
moment, we model the block by a simple interface in which data words
coming from the memory are validated by an ad-hoc signal. 
\item{{\bf cache memory}} A limited amount of on chip memory is needed
in the node. Fast access on-chip memory will be used to store control
variables (i.e. loop counters) and memory pointers. These variables were
stored in the data memory block of the control processor (T1000) in APEmille.
Indeed, these variables have very irregular access patterns and very short
access bursts. The use of relatively long-latency dynamic memory
would adversely impact performance.
On-chip memory does not need to be very large, of the order of 1K data words.
Note that, in spite of the name, this is not a true hardware controlled cache 
system, since the decision to store variables on-board or otherwise is statically 
made at compile time (one very simple strategy would be to store on-board all
non vector integer quantities defined by a program).
\end{itemize}
In any case, as seen by the processor, the memory interface has
a word-width of 128 bits (one complex double precision number) and
provides one new word at each clock cycle in burst mode. Addressing is
done on 64-bit boundaries (so real and integer variables can be stored
efficiently).

\subsection{The Instruction Cache}

Actual LQCD simulations typically spend an extremely large fraction of
the running time in just a few critical loops. For instance, a
full-fermion hybrid Monte Carlo code spends nearly 95\% of the time in the
kernel used to compute the value of the Dirac operator on the fermion
fields. Under these conditions, an instruction-cache system should have
very large efficiency. We may exploit this feature by storing node programs
in the same memory bank as data, with obvious advantages in terms of 
pin-count, real-estate reduction and cost savings.

\begin{figure}[tbh]
\epsfig{file=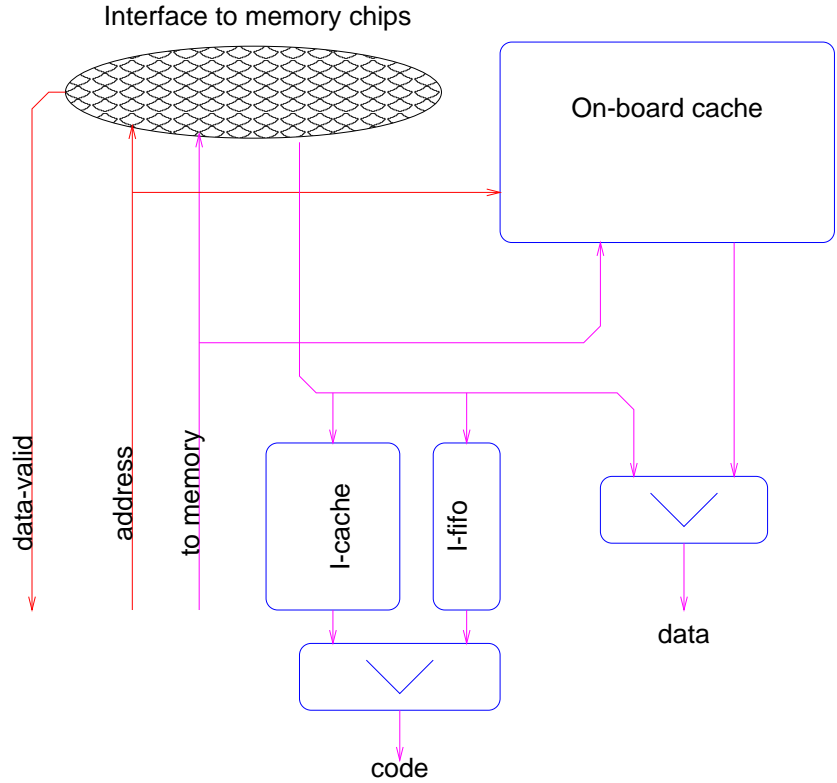}
\caption{The instruction cache and the program look-ahead system.}
\label{fig:icache}
\end{figure}

We consider a control word (micro-code word) of 128 bits, equal to
the word size that can be fetched from memory at each clock cycle.
We may modify the memory interface as shown in fig \ref{fig:icache}.

Consider for the moment just the instruction FIFO. The memory controller (not
shown in the picture)   continuously looks-ahead and prefetches instructions
from the memory, at all machine cycles in which data-memory transactions are
not in progress. Under the fully pessimistic assumption that all program cycles
involve data-memory accesses, this mechanism reduces performance by a factor 
$\leq 2$.
Now consider the instruction cache. The instruction cache is loaded
when or before the critical kernels are executed the first time
(possibly under program control: the program writer may advise the compiler
through appropriate directives that some routine or do loop is a critical
kernel).
The program then completes all following loops fetching instructions from the
cache without incurring in any time penalty. The expected efficiency
$\epsilon$ is
\begin{equation}
\epsilon = \frac{1}{f + 2 \times (1-f)},
\end{equation}
where $f$ is the fraction of cached program instructions.
If we expect to cache 90\% of all used instructions (a rather pessimistic value)
we may still expect 90\% program efficiency.

The size of the cache needed to accommodate the computational kernels is an
important parameter. We have analyzed several LQCD kernels used in TAO and
TAOmille physics programs and we have found that a cache size of the order of
16 kWords is large enough.

\subsection{The Register File} 
The register  file has the same architecture as in APEmille. The register
file has three read-only ports, one write-only port and one bi-directional
port. The read-only ports are used to feed data to the  mathematical processor,
while the write-only port stores data from the mathematical processor onto the
RF. The bi-directional port is used for memory access. All ports can be used at
each clock cycle (five independent addresses are needed). 

The word size of the processor is 64 bits, and complex numbers are stored as
pair of (adjacent) registers.

The depth of the register file affects the performance of the node. If
there are not enough registers available, temporary results cannot be hold
on-register. Memory bandwidth requirements increase and program
efficiency decreases. In APEmille, 512 registers (or 256 register pairs) were
used.  Table \ref{tab:codelength} lists the number of registers used by
critical LQCD kernels programmed in TAO and compiled for APEmille. As we see,
less than 256 are needed in all programs. We plan to design a register
file of the same size as APEmille (512 lines).
If we find out that  such a large system does not work at the required speed,
we know that the size can be halved without serious problems.

\subsection{The Mathematical Processor}
The computing engine contained in {\bf J\&T} performs three tasks:
\begin{enumerate}
\item it performs the floating-point (FP) (and, less frequently, integer)
arithmetic operations heavily used in any scientific code. This is of course
the most important functionality of the computing engine. All design trade-offs
must be guided by the aim of achieving highest possible sustained performance
for the most relevant tasks.
As already discussed, we will use the IEEE double precision format only.
The system will be heavily optimized for the arithmetics of complex-valued
numbers.
\item it computes first approximations of several important special functions
(as already remarked, these functionalities are called LUT operations
in APE jargon).
\item it performs all  (mostly integer) arithmetic and logic operations needed
to compute memory addresses. This task was carried out in a
separate chip in APEmille, with dedicated hardware. We plan to share just one
processor for this task and the previous one (and also for the fourth task,
described below). As shown elsewhere, the price paid by this optimization in
terms of performance is small.
\item it performs all arithmetic and logic operations supporting the evaluation
of branch conditions. All considerations made before about addressing also
apply here.
\end{enumerate}

We plan, for obvious reasons, to reuse to a large extent the logical design
and implementation of the arithmetic block used in APEmille (called FILU for
Floating - Integer - Logic Unit). This goal is most easily reached by extracting
from FILU the double precision FP data path, the integer data path and the LUT
circuitry, and building more complex operators as combinations of these
building blocks. We recall that the FP data path performs the {\bf normal} FP
operation ($d = a \times b + c$) and conversions between FP and integer
numbers, while the integer data path performs standard arithmetic and logic
operations in integer format. We call these basic data paths the FBB (Floating
Building Block), the IBB (Integer building Block) and the LBB (LUT Building
Block).

\begin{figure}[tbh]
\epsfig{file=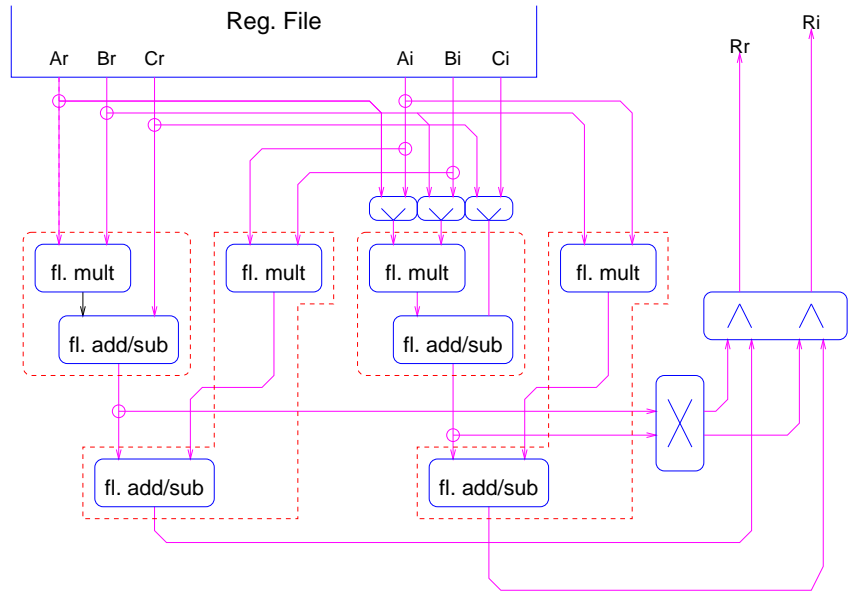}
\caption{Block diagram of the  floating point data-path (FBB)
within the mathematical processor.}
\label{fig:fpoint}

\end{figure}

Experience with the development of APEmille has taught us that a minor effort
is needed to finalize the design of the IBB and LBB. Here we consider in
details only the FBB. The architecture that we consider is shown in figure
\ref{fig:fpoint}. It uses data stored in the RF, that contains 256 register
pairs.  The two elements of the pair share the same address on each of the
three ports. A complex operand has its real and imaginary parts stored on the
same word of both registers, while a real operand sits on any location of
either block. A vector operand finally is made  up of two independent real
values, stored in the same way as a complex operand. Vector operations
can be efficiently used in LQCD codes for the generation of random numbers.

The FBB (see figure \ref{fig:fpoint}) uses four basic floating point blocks,
wired in such a way as to:
\begin{itemize}
\item compute the complex-valued version of the {\tt normal} operation:
\begin{eqnarray}
d.re = a.re \times b.re - a.im \times b.im + c.re \\
d.im = a.re \times b.im + a.im \times b.re + c.im
\end{eqnarray}
\item compute one real-valued  {\tt normal} operation on operands coming from
any element of any register.
\item compute two real-valued {\tt normal} operations on ordered operand pairs
sitting on the right and left register banks respectively (vector mode).
\end{itemize}

Some basic figures of this architecture are collected in table \ref{tab:perf}.
Control of the processor requires 5 opcode bits in the microcode word and 4 
addresses for the RF ports. In total $8 \times 4 + 5 = 37$ control bits are needed.

\begin{table}[hbt]\centering
\setlength{\tabcolsep}{.2pc}
\begin{tabular}{|l||r| |r|}
\hline\hline
type       & performance & operands in RF \\ \hline
complex    & 1600 MFlops & 256 \\
real       &  400 MFlops & 512 \\
real vect. & 800 MFlops  & 256 \\
integer    & 200 Mips    & 512 \\
int. vect. & 400 Mips    & 256 \\
\hline \hline
\end{tabular}
\vspace*{5mm}
\caption{Basic parameters of the mathematical processor.}
\label{tab:perf}
\end{table}

\subsection{Performance Estimates}
We have worked out some preliminary (but for the considered processor models
rather accurate) forecasts of the
expected efficiency of the processor outlined in the previous sub-sections on
a few computationally intensive kernels. Our results are shown in table
\ref{tab:shake} for two versions of the kernel of the Dirac operator and
for the main kernel of the Lattice Boltzmann Equation (LBE) solver.

\begin{table}[hbt]\centering
\setlength{\tabcolsep}{.2pc}
\begin{tabular}{|l||c|c|c|}
\hline\hline
Kernel     & normals & APEmille & {\bf apeNEXT} \\ \hline
Dirac1     &  336    & 44\%    & 45\%   \\
Dirac2     &  336    & 58\%    & 72\%   \\
LBE        &  870    & 55\%    & 60\%    \\
\hline \hline
\end{tabular}
\vspace*{5mm}
\caption{Measured performance on APEmille and estimated
performance of J\&T on some critical kernels
described in the text.}
\label{tab:shake}
\end{table}
 
The first two codes are appropriate for LQCD programs, while the last kernel
has been used for the simulation of turbulent fluid flows on APE systems.
The two Dirac kernels refer to a simple program written in TAO (Dirac1)
and to an highly optimized code written in assembly (Dirac2).
We can safely conclude that the performance will be in most cases
comparable or superior to APEmille.

We are still working to make our prediction more accurate and to 
test the efficiency of the processor on a larger set of computational
kernels.
   
\subsection{Implementation Issues}

In table \ref{tab:gates} we estimate the gate count of the largest
logical blocks used in J\&T. In the table we foresee a $0.18 \mu$
CMOS technology, as discussed in a previous section. Most values are evaluated
by using appropriately scaled corresponding figures for APEmille and allowing
large safety margins.
 
\begin{table}[hbt]\centering
\setlength{\tabcolsep}{.2pc}
\begin{tabular}{|l||c|c|c|c|}
\hline\hline
what       & APEmille  & scale Factor & {\bf apeNEXT} (gates) & {\bf apeNEXT}($mm^2$)\\
\hline
Reg File     & 200 K     & 2            & 400K           & 5.0\\
Math. Unit   & 100 K     & 2.5          & 250K           & 3.7\\
Intf.        &  30 K     & 2            & 60K            & 1.0\\
Data cache   &  0        & NA           & $1K \times 128 b$ & 4.4\\
Instr. cache &  0        & NA           & $16K \times 128 b$ & 34 \\
Total        & 330 K     & 4.5 + cache  & 700K + caches     & 48\\
\hline \hline
\end{tabular}
\vspace*{5mm}
\caption{Gate count and area estimate for the main components of the {\bf apeNEXT}
custom processor}
\label{tab:gates}
\end{table}

Power dissipation for this system is less than 2.5 W at 200 MHz (assuming that
about 30\% of the gates switch at each clock cycle). This processor fits
into a reasonably small die and has a relatively small pin count.

\section{The Interconnection Network}
The interconnection network is specifically tailored to the needs of LQCD
simulations. The networks supports rigid data transfers between:
\begin{itemize}
\item nearest neighbour nodes in the positive and negative direction
of the three axis (single hops)
\item next to nearest nodes, whose node-coordinate differ by $\pm$1 in two
of the three dimensions (double hops)
\end{itemize}
More formally, the network performs rigid shifts of the mesh of processors
onto itself:
\begin{equation}
(x,y,z) \rightarrow (x+ \Delta x, y +\Delta y, z + \Delta z)
\end{equation}
where $(x,y,z)$ labels the coordinates of each processor. The shifts
$(\Delta x, \Delta y, \Delta z)$, with
$|\Delta i| = 0, 1$ and $\sum_i |\Delta i| \le 2$,
are the same for each processor.

Each link has a target bandwidth of at least 300 MBytes/sec per link.
Each node needs six links to support all the above described communication
patterns.

From the point of view of system architecture the network is logically
synchronous and supports SIMD program flows, although at the layer of the
physical link, no (wall-clock) time synchronization is needed.

This definition can be made more precise in the following way:
\begin{itemize}
\item Consider a SIMD program started on all nodes of the machine. Each
node, while executing the program, starts a well defined sequence
of remote communications. The sequence is the same for all nodes.
\item we tag all remote communications by the following set of attributes:
\begin{equation}
(\Delta x, \Delta y, \Delta z, S, N)
\end{equation}
where the $\Delta$'s were defined before, $S$ is the size of the data packet
associated to the communication and $N$ is an identifier that labels all
communications issued by each program (in the following, we call
$N$ the message-tag). $N$ is initialized at 0, when starting
the program and is incremented every time a new communication is started. 
In other words, $N$ defines an ordering of all communications inside the program.
Note that all attributes of each remote communication are equal on all nodes.
\item
The network interface of each node accepts data bound to a remote 
node and tries to send it to destination. Note that although all
nodes necessarily send the same sequence of packets, the (wall-clock)
time at which a new data transfer starts may differ slightly among nodes.
The following simple protocol controls the ensuing traffic:
\begin{enumerate}
\item Each network interface refuses to accept a data packet coming from
a different node and tagged by $N$ unless it has been already instructed
by its own node to start transferring $N$.
\item Each network interface delivers incoming data in strict ascending
$N$ order.
\end{enumerate}
This protocol is needed to make sure that all messages reach 
destination in the appropriate ordering. As we see,
very simple rules are needed to
reach this goal under the assumption that programs follow the SIMD paradigm.
\end{itemize}

\begin{figure}[tbh]
\epsfig{file=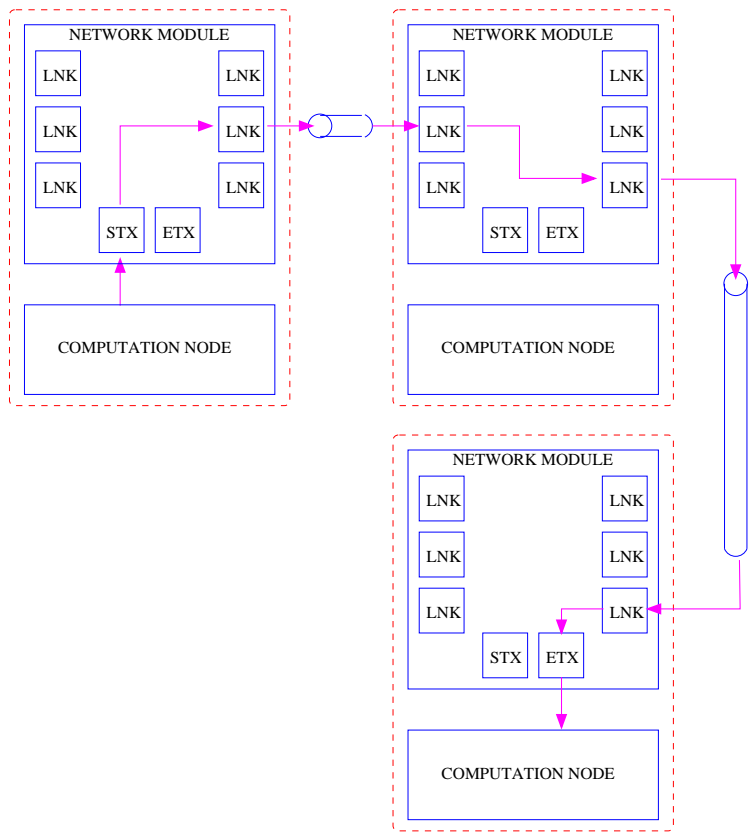}
\caption{Architectural block diagram of the interconnection network.}
\label{fig:network}
\end{figure}

Note that the network can perform several useful sanity checks:
\begin{itemize}
\item Once a node instructs the network to send a data-packet to a given
destination, the network implicitly knows which packets it should expect
on its links with a given tag. For instance, if a data-packet must be routed
to {\tt south - east}, then data with the same tag is expected from 
{\tt west} for delivery at the local node and from {\tt north} to be
routed to {\tt east}. The network interface can check that this is actually
the case.
\item The network can also check whether the right sequence of tags is
received within a (programmable) time-out delay.
\item The network can further check whether the data-sizes of all messages
associated to a given tag are equal.
\end{itemize}
All these checks are important to help debug either ill-functioning hardware or
wrong programs.

Error rates in the network are an important issue.
In plain fact, we do not know the Bit Error Rate (BERR)
that we may expected on fast links.
We are only able to quote the upper limit on the BERR
implied by the safe operation of the APEmille prototype
($10^{-15}$).
The BERR value needed for an error free {\bf apeNEXT} operation on runs lasting
a  few days  is an embarrassingly low $10^{-17}$.
For this reason, we have decided to stay on the safe side and
to design a network partially able to
recover from errors. If we are able to recover from errors, a much
more manageable picture emerges: for instance a 
more modest BERR $\simeq 10^{-12}$
implies the failure of one transmission burst every second on the entire
machine.

\begin{figure}[tbh]
\epsfig{file=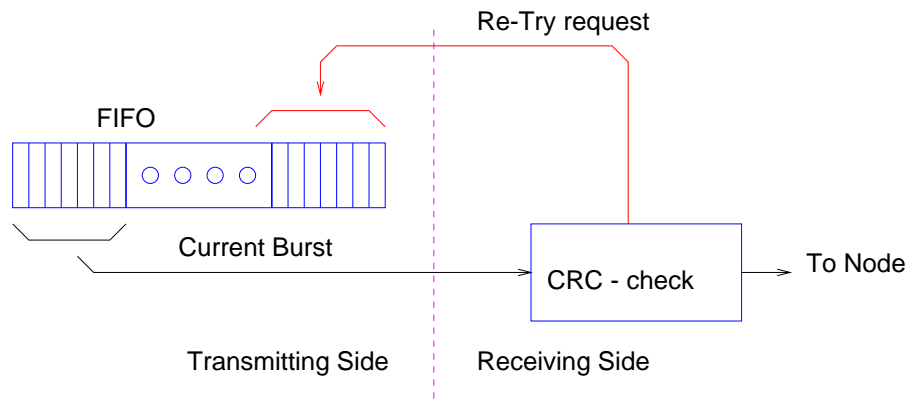,width=12cm}
\caption{Low-latency retry mechanism for the {\bf apeNEXT} links.}
\label{fig:error}
\end{figure}

We are considering a retry mechanism, shown in figure \ref{fig:error} that has
only a moderate impact on communication latency:
\begin{itemize}
\item We divide each data block traveling on a remote link in relatively
short bursts (say, 16 bytes) followed by a cyclic-redundancy-check (CRC).
\item Data bursts are sent from the transmitting nodes, followed
by their CRC. A small number of bursts is kept on
the transmitting node, stored inside a FIFO queue, also after transmission.
\item The receiving end of the link checks the CRC of each burst as
it arrives. If the check is successful it delivers received data. The
latency implied by this procedure is not longer than the size of each burst.
\item If an error is detected, the receiving end requests the corrupted 
burst to be retransmitted. This is possible, since relevant data
is still available on the transmitting side of the link.
\end{itemize}

We plan to finalize most of the high level details of the network using a black
box model of the physical link layer. In this way we can complete most
of the design even before selecting the actual link technology. In parallel,
we are already carrying out tests on some of the link technologies.

\section{Machine Assembly and Partitioning}
We plan to assemble a certain number of APEmille processors on a
printed circuit board (PCB). Preliminary evaluations suggest that 16 processors
can be  placed on one PCB, of roughly the same size as the one used
for APEmille. For comparison, note that one APEmille PCB houses 8
processors. In this case however a large (almost 50\%) fraction of the real
estate is used by the control processor and ancillary circuitry.
If we use PCB's of the same size as APEmille, we can immediately reuse
the mechanical components of the older system.

At this point in time we have two options for the topology of the nodes
belonging to one PCB. The first option is a three-dimensional structure
with $2 \times 2 \times 4$ processors. The second option implies a
two-dimensional set-up of $4 \times 4 $ processors. PCB's are assembled
inside a crate. All PCB's inside one crate are connected to a communication
backplane. If we use the mechanical components developed for APEmille, we can
reasonably house up to 16 PCB's inside one crate. Larger machines
use more crates.

If we use the first option for the node topology inside a PCB, we can allocate
onto the backplane all links in two of the three dimensions (say, directions y
and z), building a system of size $4 \times 8 \times 8$. Communications
in the x direction are implemented via cable links. If we assemble
and connect together $n$ crates, we obtain {\bf apeNEXT} systems of
size $(4\times n) \times 8 \times 8)$. This option is very similar to the one
used by APEmille, where systems of size $(2\times n) \times 8 \times 8$ are
allowed. We call this arrangement Option 1A.

It is possible to use the same structure of the PCB as above, wiring however
the unit inside a crate according to a $4 \times 4 \times 16$ topology.
In this case large machines contain
$ (4 \times n) \times (4 \times m) \times 16 $ nodes. This is option 1B.

In the case that the second option for the PCB
is selected, we allocate all links belonging
to one of the spatial directions (say, direction z) onto the backplane.
Links in the x and y directions stemming out of the PCB use cable links instead.
Using this arrangement, systems of size
$ (4 \times n) \times (4 \times m) \times 16 $ can be assembled.
This is option 2.
 Some basic
figures relevant to both options are collected in table \ref{tab:mecca}.

\begin{table}[hbt]\centering
\setlength{\tabcolsep}{.2pc}
\begin{tabular}{|l||c|c|c|}
\hline\hline
        & Option 1A & Option 1B  & Option 2 \\
\hline
number of procs.    &  16   &16 & 16 \\
Peak PCB perf. & 25.6 GFlops & 25.6 GFlops& 25.6 GFlops\\
topology            & $2 \times 2 \times 4$ & $2 \times 2 \times 4$ & 
                      $1 \times 4 \times 4$ \\
crate topology      & $2 \times 8 \times 8$ & $ 4 \times 4 \times 16$ &
                      $ 4 \times 4 \times 16$ \\
Large-systems & $ (2 \times n) \times 8 \times 8$ &
                      $ (4 \times n) \times (4 \times m) \times 16$ &
		      $ (4 \times n) \times (4 \times m) \times 16$\\
Remote links (PCB)  & 40         &40          & 48 \\
Remote links (BP) & 32           &24          & 32 \\
Remote links (cables)    & 8     &16          & 16 \\
\hline \hline
\end{tabular}
\vspace*{5mm}
\caption{Basic figures of three possible {\bf apeNEXT} machine configurations.}
\label{tab:mecca}
\end{table}

In both cases, a large number of signal must be routed on the backplane. This
is a serious but not formidable engineering problem. Assuming that 20 data lines
are needed per link, we have 640 pins carrying data from the PCB to the
backplane (see again table \ref{tab:mecca}). This requires as little as about
17 cm on the PCB edge, using high-density high-speed matched-impedance
connectors developed by several vendors (see for instance \cite{teradyne}).
Of course, special care must be taken in the design of accurately matched
transmission lines, both on the backplane and on the main PCB. 

As discussed earlier, the backplane must also house a CPCI backplane. This is
made by a straightforward copy the well tested CPCI backplane developed for
APEmille.

We will decide later on in the design phase on the selected topology, using
information from test setups and taking also into consideration the relative
merits of the two solutions from the point of view of physics simulations.

\section{Software: The Programming Environment}
The {\bf apeNEXT} programming environment will be initially based on two main lines:
\begin{itemize}
\item The TAO programming language, extensively used in APE100 and in APEmille,
will be supported. This is necessary to allow easy and early migration of the
large set of existing QCD programs on the new machine This large
portfolio of programs is also going to be extremely useful for test and
debugging purposes.

We do not plan to make any substantial improvement to TAO. We will just modify
the back-end section of the TAO compiler, so it produces {\bf apeNEXT} assembly
codes.

\item 

We plan to develop a C/C++ language compiler for {\bf apeNEXT} since the early phases
of the project. 
Very few extensions will be added to the standard C syntax,
with the goal of minimizing the effort for the programmer in learning a new
language. SPMD parallelism will be realized by just a few special constructs,
similar to the ones already present in TAO:

\begin{enumerate}
\item the {\tt where} statement executes code blocks based on
local conditions.
\item the {\tt all, none, any} keywords in
a standard C condition perform aggregate evaluation of local conditions.
\item Remote communications will be specified by constant pointers.
\end{enumerate}

The compiler will be implemented by porting already available public domain
compilers (like, for instance, the GNU C/C++ compiler or {\it lcc}) with the 
needed SPMD extensions in the front-end (the language definition) and all 
necessary changes in the back-end, to produce the target assembly.
A similar GNU-gcc based compiler prototype for APEmille \cite{wwwbetto}
is currently evaluated.

Note that, by using already available compilers, it will be relatively easy to
implement all SPMD extensions of the language on more traditional machines
(like PC clusters).  Conversely, already developed parallel programs written in
C (and following the SIMD/SPMD paradigm) will be easily ported onto {\bf apeNEXT}. We
regard this possibility as our main path to build a common programming
environment between {\bf apeNEXT} and more traditional systems.
\end{itemize}

\begin{figure}[tbh]
\epsfig{file=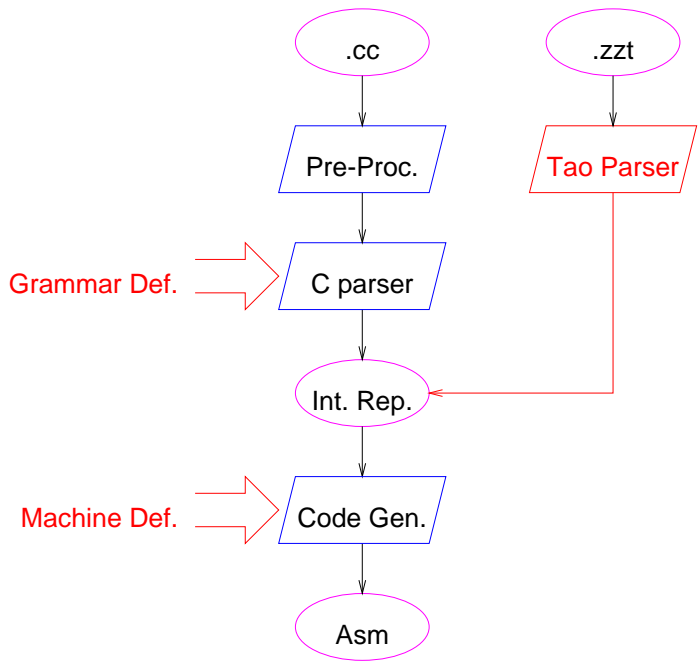}
\caption{A sketchy view of the internal
structure of the GNU compiler, including
planned extensions for the {\bf apeNEXT} software environment.}
\label{fig:software}
\end{figure}

We want to push still further the goal of a more general {\bf apeNEXT} programming
environment.
We plan to merge to some extent the programming environments based on Tao
and on C/C++, and at the same time enhance the portability of programs between
APE systems and more traditional computer clusters. We plan to work according
to the lines described in figure \ref{fig:software}, that uses for
definiteness the structure of the GNU compiler.

The blue boxes in figure \ref{fig:software} sketchily
describe the overall organization
of a modern compiler. There is a front-end block with
a configurable parser that transforms the user code into an internal 
representation, based on a tree representation of the code and symbol
tables. The back-end block maps the internal representation onto assembly 
code for a specific target architecture. 

The APE C compiler can be implemented on the basis of existing and configurable
front-ends with minor modifications to include the required syntax extensions 
for parallel processing. The back-end section must of course be customized
to produce {\bf apeNEXT} assembly code.

It is also possible to add an additional parser at the front-end
level. (For instance, Fortran is implemented in this way in the GNU compilers.)
We intend to follow this path and to include the TAO parser,
suitably modified to generate the GNU internal representation.
Indeed, TAO cannot be easily handled
by standard configurable parsers because
of its dynamic grammar. 
In figure \ref{fig:software} the APE specific extensions are drawn in red.

When the program outlined above is accomplished, we will
have a very neat portable environment in which:
\begin{itemize}
\item all powerful optimization techniques of a standard compiler core 
      are available.
\item TAO and C codes can be compiled for a standard computer system
(e.g. a PC).
\item C and TAO codes can be compiled for an APE system.
\end{itemize}

The design and implementation of this open programming environment is
a long term and very high priority goal of our project. This is not
going to be an easy task and can certainly not be finalized in short terms.
Physics exploitation of {\bf apeNEXT} in the early
phases does not depend on this environment, since the traditional 
APE software tools can be used.
 
At the machine level, we will port to the new architecture and improve the
well-established VLIW code-scheduling and code-compressing tools already used
in APEmille. 

\section{Software: The Operating System}
We plan to shape the {\bf apeNEXT} operating system as a direct evolution of the
APEmille system: the basic idea is that we use as much as possible the
services provided by Linux on the network of host PC's.
\begin{itemize}

\item The {\bf apeNEXT} operating system must load executable codes on the array
of processing nodes and start execution of the whole system. We remind
that all steps of program compilation and optimization are not done on
the {\bf apeNEXT} processors itself but are performed on the host PC's (or on
any other Linux machine).

\item The second task performed by the operating system is the support
for input-output operations requested by the executing program.
Note here that these operations use the standard file systems available on
the host PC's (or, on any networked disk server). Of course, large
data transfers, where high bandwidth is needed, are performed in parallel 
by all PC's on local disks. 
Later on, we will make some additional remarks on this point.

\item The third task performed by the operating system is the
monitoring and control of all nodes at a low level. Typical examples
include the inspection and setting of status-registers, the analysis
of error conditions, explicit writes or reads to memory locations.
\end{itemize}

All functions described above are handled by the APEmille operating system
in a reasonably efficient and user-friendly way. Most operations
can be easily moved onto {\bf apeNEXT} by rewriting only
the lowest layer levels of the operating system, like device drivers
or the functions mapping a specific operation on a specific node
onto the appropriate PC. We expect therefore to be able to put to work
quickly an early version of the system.

An area where we would like to come up with new ideas, not needed however
for the early commissioning of {\bf apeNEXT},  is some
version of a parallel file system, where large field and propagator
configurations can be stored in a standard format. This is an obvious
starting point to allow the sharing of QCD configurations among
collaborating groups.
We see this work as a partial contribution of {\bf apeNEXT} to the GRID project.
Work on this line will be therefore coordinated with GRID.
 
\section{Design Method}
In designing the needed VLSI components as well as the overall system,
we want to follow the method used in APE100 and APE1000, with a
number of improvements  to make it more efficient and faster. The main
advantage of this method has been shown in APE100 and
APEmille: in both cases all components of the machine were designed
"first-time-right". 

The main idea behind our method is some informal implementation
of "hardware-software co-design" techniques: 

\begin{itemize}
\item We base our design on a VHDL model of a large and significant
fraction of the whole system. The model contains all in-house  developed
systems as well as all off-the-shelf components. Initially, the model
will be a very crude approximation to the actual system, gradually
incorporating all details. This reference model is available at all 
collaboration sites.
\item All VLSI (or FPGA based) components of the system are derived
with high quality synthesis tools from the VHDL design. In the (hopefully
rare) cases where some components cannot be synthesized from a VHDL description,
a VHDL model is built anyway, and test vectors for the actual implementation
are derived by the VHDL model. Non-VLSI parts of the systems (i.e.,
processing boards) will be modelled in VHDL by their designers.
\item The VHDL model supports a reasonable approximation of the interaction with
the host system (operating system).
\item All software developments are immediately tested on the VHDL model.
At an early stage of the design, tests will involve performance estimates on
crudely modelled architectural choices. As the level of details of the
model increases, actual programs, in all their intricacies, will be executed
on the model, giving quick feed-back on any design detail.
\end{itemize}

As an improvement with respect to APEmille, we will insist on:
\begin{itemize}
\item a continuous availability of the model at all collaboration sites, so
all members of the collaboration can easily monitor the effects of a design
change made elsewhere. This can be achieved with reasonable effort by keeping
a master copy of the model on an AFS cell available from all sites.
\item an effort to allow access to the model from a basic version of the
operating system, so that even the more physics-oriented members of the
collaboration can exercise it.
\item a systematic use of blind-test procedures: at all stages of the design
phase, test sequences for any part of the machine will be prepared and
executed by someone who has not been directly involved in the design.
\end{itemize}

Finally, we address the issue of the design of some VLSI blocks
that depend critically on some element which is either not under
our direct control or not completely defined at this stage.
Examples of this situation are the details of the memory system
(which heavily depends on the type of memory available at the time when the
prototypes are built) for the custom processor, or the actual choice
for the physical layer of the interconnection links. 
Changes made in memory technology during the design lifetime of
APEmille have indeed adversely affected that project, as large subsystems
within  J1000 and T1000 processors had to be modified to adapt to changing
memory specifications.

We want to solve this problem by confining all details of the memory
and link interfaces in a "grey box", that interacts with the rest of the design
with some simple and rather general data and control paths. In this way,
a very large fraction of the design can be finalized independently,
while the interface-specific blocks will be procured (if available from
external sources) or
designed in detail at the last moment.

\section{Conclusions}
This document has described physics requirements and the basic architecture
of a next generation LQCD computer project. We think that
the well-tested SIMD/SPMD architecture of the
previous APE generations is still the best choice for a LQCD-focused 
high performance engine. At the engineering level,
we plan to use technologies similar to those used in APEmille.
We think that this choice reduces development costs and risks.

In the near future, collaboration between groups 
active in LQCD simulations will become tighter and tighter. For this reason we
plan to work hard on the development of a software environment
allowing easy migration between {\bf apeNEXT} and more traditional computers.  

\section*{Acknowledgments}

Several people have helped shape the basic ideas of {\bf apeNEXT} and
contributed several important pieces of information. We would like to thank I.
D' Auria, W. Friebel, J. Heitger, M. Loukyanov, S. Menschikov,  A. Sapojnikov,
T. Sapojnikova, L. Schur, A. Thimm. N. Christ, N. Eicker, Y.
Iwasaki, T. Lippert and K. Schilling have provided valuable information on the
status and perspectives of their projects.

\eject

\end{document}